\documentclass[aps, prb, 10pt, twocolumn, showpacs, showkeys, floatfix]{revtex4-1}

\usepackage{amsmath, amssymb, mathtools, graphicx, array, hyperref, float}

\newcommand{\nn}				{\nonumber}

\newcommand{ \ket }			[1] { \left|{#1}\right> }

\newcommand{ \ex }				[1] { \left<{#1}\right> }

\newcommand{\w}				{\omega}

\newcommand{\W}				{\Omega}

\newcommand{\ft}				{\mathcal{T}}
\newcommand{\fg}				{\mathcal{G}}

\newcommand{\fr}				{\mathcal{R}}
\newcommand{\fm}				{\mathcal{M}}
\newcommand{\fq}				{\mathcal{Q}}
\DeclareMathOperator{\tr}{tr}

\begin{document}


\title{Momentum and position detection in nanoelectromechanical systems beyond Born and Markov approximations}
\author{Stefan Walter}
\author{Bj\"orn Trauzettel}
\affiliation{Institute for Theoretical Physics and Astrophysics, University of W\"urzburg, 97074 W\"urzburg, Germany}
\date{\today}

\begin{abstract}
We propose and analyze different schemes to probe the quantum nature of nanoelectromechanical systems (NEMS) by a tunnel junction detector. Using the Keldysh technique, we are able to investigate the dynamics of the combined system for an arbitrary ratio of $eV/\hbar \Omega$, where $V$ is the applied bias of the tunnel junction and $\Omega$ the eigenfrequency of the oscillator. In this sense, we go beyond the Markov approximation of previous works where these parameters were restricted to the regime $eV/\hbar \Omega \gg 1$. Furthermore, we also go beyond the Born approximation because we calculate the finite frequency current noise of the tunnel junction up to fourth order in the tunneling amplitudes.

Interestingly, we discover different ways to probe both position and momentum properties of NEMS. On the one hand, for a non-stationary oscillator, we find a complex finite frequency noise of the tunnel junction. By analyzing the real and the imaginary part of this noise separately, we conclude that a simple tunnel junction detector can probe both position- and momentum-based observables of the non-stationary oscillator. On the other hand, for a stationary oscillator, a more complicated setup based on an Aharonov-Bohm-loop tunnel junction detector is needed. It still allows us to extract position and momentum information of the oscillator. For this type of detector, we analyze for the first time what happens if the energy scales $eV$, $\hbar \Omega$, and $k_B T$ take arbitrary values with respect to each other where $T$ is the temperature of an external heat bath. Under these circumstances, we show that it is possible to uniquely identify the quantum state of the oscillator by a finite frequency noise measurement.
\end{abstract}


\maketitle

\section{Introduction}

Nanoelectromechanical systems (NEMS) have become a promising playground for probing the quantum behavior of
mesoscopic objects, theoretically as well as experimentally \cite{Schwab:2005p299,Clerk:2010p109}. The diverse reasons to study NEMS
are their vast number of (possible) applications as for instance the measurement of mass, force and position \cite{
Yang:2006p305,Mamin:2001p304,LaHaye:2004p303,Lassagne:2009p310,Steele:2009p312} with high precision.

Nanomechanical systems being at the boarder of classical to quantum are also being studied from a very fundamental
point of view. This includes the observation, measurement and control of quantum states of a mesoscopic mechanical
continuous variable system such as a harmonic oscillator. Superconducting qubits electrically coupled to the mechanical
system have been successfully used to characterize the mechanical resonator's quantum state \cite{Hofheinz:2009p308}.

Making quantum effects visible in nanomechanical systems calls for ultralow temperatures and low dissipation. The goal
of observing the quantum mechanical ground state of a harmonic oscillator requires temperatures $k_{B} T \ll \hbar \Omega$.
Recently, this goal has been achieved by using a microwave-frequency mechanical oscillator, with a frequency of $f \approx 6 \, GHz$
which allowed cooling to the ground state with conventional cryogenic refrigeration \cite{OConnell:2010p43}. Further proposals
of cooling a nanomechanical resonator coupled to an optical cavity have been proposed \cite{Marquardt:2007p48,WilsonRae:2007p7}
and experimentally implemented \cite{Rocheleau:2010p309,Naik:2006p122}.

The theoretical treatment of NEMS widely uses a Markovian master equation approach \cite{Clerk:2004p185,Doiron:2007p5,
Doiron:2008p13} with a few exceptions, for instance, the work by Wabnig {\it et al.} \cite{Wabnig:2007p16} and Rastelli {\it et al.}
\cite{Rastelli:2010p315} where a Keldysh perturbation theory has been employed. Here, we also make use of the Keldysh
technique because it allows us to treat the non-equilibrium system fully quantum mechanically and, furthermore, to carefully
investigate the non-Markovian regime where $eV \ll \hbar \Omega$. Since we are interested in the quantum nature of the oscillator,
it is important that the temperature $T$ and the applied bias $V$ of the tunnel junction are not much larger than the eigenfrequency
$\Omega$ of the oscillator. Otherwise, the oscillator would be heated and low energy properties inaccessible.

The article is organized as follows. Our key results are summarized in Sec.~\ref{sec:keyresults}.
In Sec.~\ref{sec:model}, we introduce the generic model.
This is followed by an introduction of the formalism we use in Sec.~\ref{sec:formal} with subsections focusing on the fermionic
reservoir and on the oscillator dynamics.
The main part of this article is presented in Sec.~\ref{sec:perturb}, where we discuss the calculation as well as the results for the finite frequency current noise. Finally, we conclude in Sec.~\ref{sec:conclusion}.

\section{Key results of the paper}
\label{sec:keyresults}

The motivation of our work is to study an experimentally feasible setup in which the quantum nature of NEMS can be probed by current noise measurements of a tunnel junction detector. A quantum NEMS can be described by a quantum harmonic oscillator which is a continuous variable system characterized by two non-commuting operators $\hat{x}$ and $\hat{p}$. Therefore, it is desirable to have a detector at hand that can measure expectation values with respect to $\hat{x}$-dependent observables, $\hat{p}$-dependent observables, as well as observables that depend on both $\hat{x}$ and $\hat{p}$.

\begin{figure}[ht]
	\center{\includegraphics[width=0.9\columnwidth]{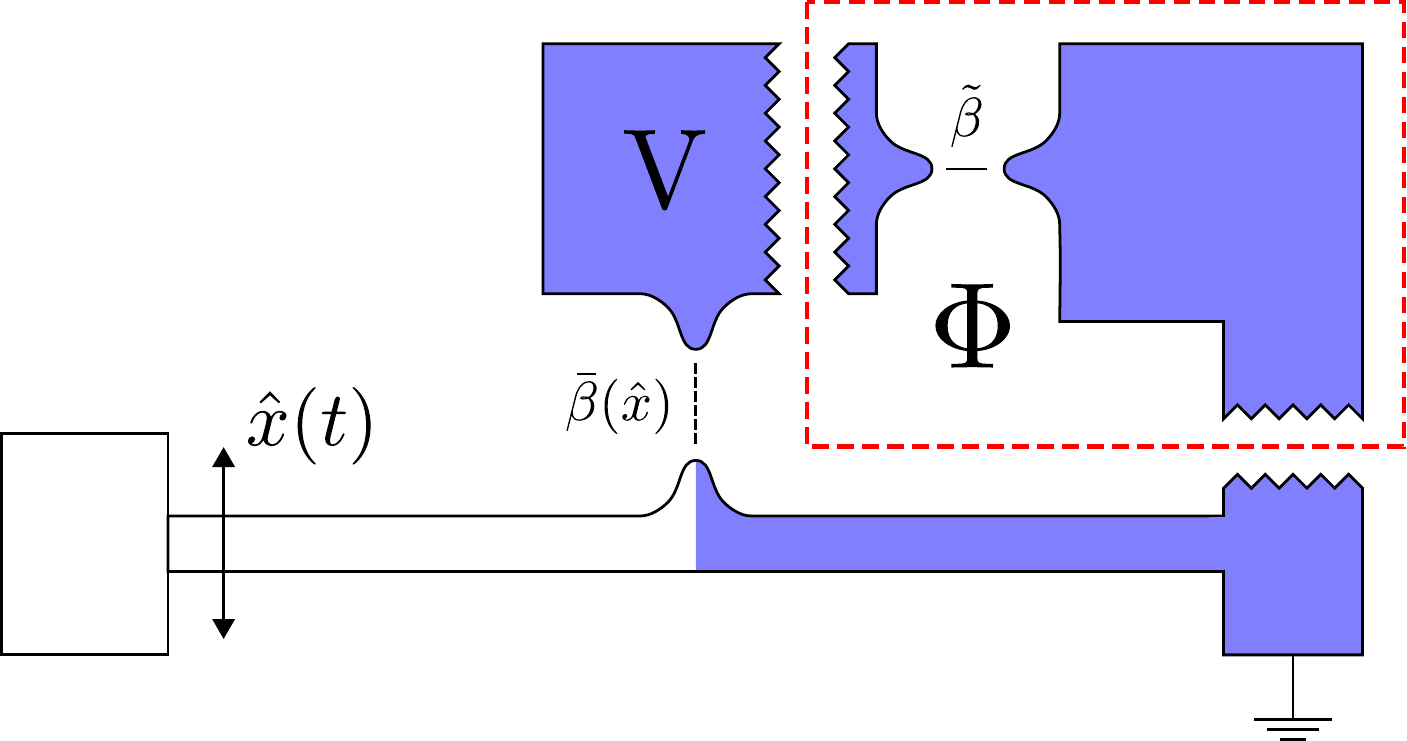}}
	\caption{ Schematic setup for the realization of a position detector which can be extended by the parts enclosed
			in the red dashed box to a momentum detector. The total tunnel amplitude in the case of the position
			detector is given as $\beta = \bar{\beta}(\hat{x}) = t_{0} + t_{1} \hat{x}$ and in the case of the momentum detector
			as $\beta = \bar{\beta}(\hat{x}) + \tilde{\beta} = t_{0} + t_{1} e^{i \eta} \hat{x}$, where the relative phase $\eta$
			between $t_{0}$ and $t_{1}$ can be tuned via a magnetic flux $\Phi$ penetrating the AB loop. If the oscillator
			is in a non-stationary state, already the parts without the elements in the red dashed box serve as a position
			as well as a momentum detector. }
	\label{fig:schemSetup}
\end{figure}

In Ref.~\onlinecite{Doiron:2008p13}, Doiron {\it et al.} have proposed a setup which could be used for position and momentum detection of NEMS. This setup consists of two tunnel junctions forming an Aharonov-Bohm (AB) loop. There, it is possible to tune the relative phase between the tunnel amplitudes (where one depends on $\hat{x}$ and the other one not) via a magnetic flux penetrating the AB loop, see Fig.~\ref{fig:schemSetup} for the schematic setup. In such a setup, the symmetrized current noise
\begin{equation} \label{ssym}
S_{\rm sym}(\omega) = \frac{1}{2} \int dt e^{i \omega t} \left\langle \left\{ \Delta \hat{I}(t) , \Delta \hat{I}(0) \right\} \right\rangle
\end{equation}
(with the current fluctuation operator $\Delta \hat{I}(t) = \hat{I}(t) - \langle \hat{I} \rangle$) of the tunnel junction detector can either contain information on the
oscillator's position spectrum
\begin{equation} \label{sx}
S_{x}(\omega) = \frac{1}{2} \int dt e^{i \omega t} \left\langle \left\{ \hat{x}(t) , \hat{x}(0) \right\} \right\rangle,
\end{equation}
then $S_{\rm sym}(\omega) \sim S_{x}(\omega)$, or the oscillator's momentum spectrum
\begin{equation} \label{sp}
S_{p}(\omega) = \frac{1}{2} \int dt e^{i \omega t} \left\langle \left\{ \hat{p}(t) , \hat{p}(0) \right\} \right\rangle,
\end{equation}
then $S_{\rm sym}(\omega) \sim S_{p}(\omega)$, see also Eq.~(\ref{eqn:m1}) below. The current noise $S_{\rm sym}(\omega)$ can also
contain information on both, the position and the momentum of the oscillator. The former case has been coined $x$-detector
and the latter case $p$-detector. We note that $S_{\rm sym}(\omega)$, $S_{x}(\omega)$, and $S_{p}(\omega)$ are properly defined
above in Eqs.~(\ref{ssym})-(\ref{sp}) for a stationary problem. In the non-stationary case, which is also subject of discussion in this
work, these quantities do not only depend on a single frequency $\omega$ but on one frequency argument and one time argument
instead, see Eq.~(\ref{eqn:g7}).

To be more specific, in this article, we call an $x$-detector, a detector that allows to measure
expectation values of the oscillator's position operator $\hat{x}$, i.e. $\ex{\hat{x}}$, $\ex{\hat{x}\hat{x}}$, etc.. Similarly, we call a
$p$-detector, a detector that allows to measure expectation values of $\hat{p}$, the oscillator's momentum operator, i.e.
$\ex{\hat{p}}$, $\ex{\hat{p}\hat{p}}$, etc.. In Ref.~\onlinecite{Doiron:2008p13}, switching from the $x$-detector to the $p$-detector
is then accomplished by tuning the relative phase between the tunnel amplitudes. The main difficulty of this setup is the
need of long coherence times and length in the AB loop to make the switching possible. Here, we show that the AB setup can be
avoided. We find that the current noise of the coupled oscillator-junction system with one tunnel junction only, already can
be used for momentum detection due to the complex nature of the current noise when the oscillator is in a {\it non-stationary} state.
This is the first key result of our work, specified and intensively discussed in Sec.~\ref{sec:meas} below.

We further investigate the current noise stemming from a stationary oscillator up to fourth order in the tunneling amplitudes, thereby going beyond the Born approximation. Most importantly, we extend previous results of Ref.~\onlinecite{Doiron:2008p13} to the non-Markovian regime without any restrictions on the relative magnitude of the energy scales $eV$, $\hbar \Omega$, and $k_B T$. We show that peaks in the finite frequency current noise at $\omega = \pm \Omega$ (both for the $x$-detector and the $p$-detector) are a fourth order effect. In the Markovian regime, the peaks in the position detector signal are always much larger than the ones in the momentum detector signal. This is different in the non-Markovian
regime. There, we even find a larger signal for the momentum detector compared to the position detector, clearly demonstrating that the non-Markovian regime is the preferred regime to operate the momentum detector. The detailed understanding of the $x$-detector and the $p$-detector developed in this article allows us to uniquely identify the quantum state of the oscillator by a finite frequency noise measurement. This is the second key result of our work, specified and intensively discussed in Sec.~\ref{sec_pdet} and Sec.~\ref{sec_number} below.

%

\section{Model}
\label{sec:model}

The system we consider consists of a nanomechanical harmonic oscillator coupled to a biased tunnel junction. In
Ref.~\onlinecite{FlowersJacobs:2007p188} an experimental realization is shown, where electrons can tunnel from an
atomic point contact (APC) onto a conducting oscillator. The coupled system is described by the following Hamiltonian
\begin{equation}\label{eqn:a1}
	\hat{H} = \hat{H}_{\rm osc} + \hat{H}_{\rm res} + \hat{H}_{\rm tun} \, ,
\end{equation}
with
\begin{align}
	\hat{H}_{\rm osc}	&= \frac{\hat{p}^{2}}{2m} + \frac{1}{2}m\Omega^{2}\hat{x}^{2} \label{eqn:a2a} \\
	\hat{H}_{\rm res}	&= \sum_{l,r}\varepsilon_{l} \, \hat{c}^{\dag}_{l} \, \hat{c}_{l} + \varepsilon_{r} \, \hat{c}^{\dag}_{r} \, \hat{c}_{r} \label{eqn:a2b} \, ,
\end{align}
where $\hat{H}_{\rm osc}$ describes the oscillator with $\hat{x}$ and $\hat{p}$ being the position and momentum operator of
the oscillator with mass $m$ and frequency $\Omega$, respectively. $\hat{H}_{\rm res}$ contains the fermionic reservoirs of
the left and right contacts. The oscillator couples to the tunnel junction via the tunneling Hamiltonian
\begin{equation}\label{eqn:a3}
	\hat{H}_{\rm tun} = \sum_{l,r} \beta \, \hat{c}^{\dag}_{l} \, \hat{c}_{r} + h.c. \, .
\end{equation}
Here $\hat{c}_{i}$ ($\hat{c}^{\dag}_{i}$) annihilates (creates) an electron in reservoir $i=l,r$. Motivated by the experimental setup
in Ref.~\onlinecite{FlowersJacobs:2007p188}, we take the oscillator to act as one of the fermionic reservoirs. Therefore, the
tunneling gap depends on the position of the oscillator, modifying the tunneling amplitude of the APC. For small oscillator
displacements $x$, we assume linear coupling of the oscillator to the tunnel junction with a tunnel amplitude $\beta_{1}$.
Hence we obtain $\beta = \left[ \beta_{0} + \beta_{1} \,\hat{x} \right]$, with $\beta_{0}$ being the bare tunneling amplitude.
Here, we allow for complex tunnel amplitudes $\beta_{0}$ and $\beta_{1}$ as previously discussed in Refs.~\onlinecite{Doiron:2008p13, Schmidt:2010p301}.
With $\eta$ we denote the relative phase between the tunnel amplitudes, i.e. we write $\beta_{0} = t_{0}$ and $\beta_{1} = t_{1} e^{i \eta}$ where $t_{0},t_{1} \in \Re$.
A possible experimental realization of the finite and tunable phase $\eta$ is discussed in Ref.~\onlinecite{Doiron:2008p13}.
As a consequence this phase $\eta$ gives rise to the possibility to detect the oscillator's momentum expectation value
$\ex{\hat{p}^{2}}$, present in the current noise.

\section{Green's functions of the fermionic reservoir and of the oscillator using the Keldysh technique}
\label{sec:formal}

The true non-equilibrium, non-Markovian quantum behavior of the coupled system is the subject of our interest. Therefore,
we make use of the Keldysh formalism \cite{Rammer:1986p158,Rammer:2007p314} for our calculation. The quantities being accessible in the
experiment are for instance the tunnel current and the current-current correlator, the noise of the tunnel junction. These will also be
the main objects of interest in this article. We employ a perturbation theory in the tunnel Hamiltonian $\hat{H}_{\rm tun}$ and
calculate the noise up to fourth order in the tunneling.

The current operator is given by $\hat{I} = -e \dot{\hat{N}}_{l}$, where $\hat{N}_{l} = \hat{c}^{\dag}_{l} \hat{c}_{l}$ counts
electrons in the left reservoir. We then write the current operator as
\begin{equation}\label{eqn:b1}
	\hat{I} = e \left[ \hat{j}_{0} + \hat{x} \hat{j}_{1} \right]
\end{equation}
and similarly
\begin{equation}\label{eqn:b2}
	\hat{H}_{\rm tun} = \hat{h}_{0} + \hat{x} \hat{h}_{1} \, ,
\end{equation}
with $\hat{j}_{i} = i \left[ \hat{\ft}_{i} - \hat{\ft}^{\dag}_{i} \right]$ and $\hat{h}_{i} = \hat{\ft}_{i} + \hat{\ft}^{\dag}_{i}$. The operator
$\ft_{i}$ is given by $\ft_{i} = \sum_{l,r} \beta_{i} \hat{c}^{\dag}_{l} \hat{c}_{r}$ with $i \in \{ 0,1 \}$.

\subsection{Reservoirs Green's functions}
\label{sec:resGF}

\begin{figure}[ht]
	\center{\includegraphics[width=0.9\columnwidth]{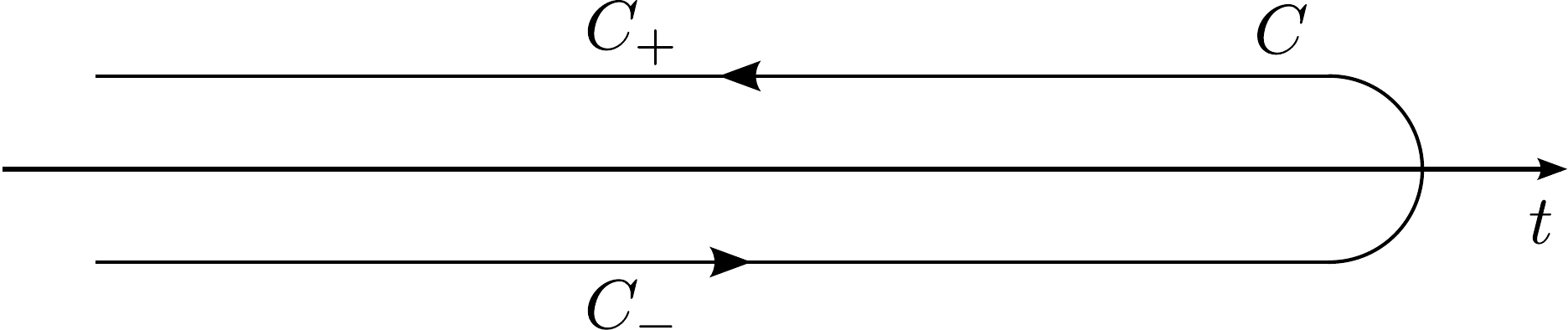}}
	\caption{ Keldysh contour $C$ with the lower branch $C_{-}$ and the upper branch $C_{+}$. }
	\label{fig:keldyshC}
\end{figure}

The fermionic Green's functions of the left and right reservoirs (free electron gas) $G_{l,r}$, are given on the Keldysh contour $C$
by $G_{l,r}(t,t') = -i \ex{T_{c} \, \hat{c}_{l,r}(t) \hat{c}_{l,r}^{\dag}(t')}$. $T_{c}$ denotes the time ordering operator on the Keldysh
contour, placing times lying further along the contour to the left. Figure~\ref{fig:keldyshC} shows the Keldysh contour $C$. The contour
consists of a lower branch, $C_{-}$ on which time evolves in forward direction and of an upper branch, $C_{+}$ where time evolves
in backward direction. Switching from times lying on the contour to real times is done by analytic continuation. The Keldysh
Green's functions $G_{l,r}(t,t')$ can then be represented by a matrix. A Fourier transformation leads to the following Green's functions
\begin{align}\label{eqn:c1}
	G_{l,r}(\omega) &=  \left(\begin{array}{cc} G_{l,r}^{--}(\omega) & G_{l,r}^{-+}(\omega) \\ G_{l,r}^{+-}(\omega) & G_{l,r}^{++}(\omega) \end{array}\right) = \\ \nn
				 &= 2\pi i \rho_{0} \left(\begin{array}{cc} n_{l,r}(\omega)-1/2 & n_{l,r}(\omega) \\ n_{l,r}(\omega)-1 & n_{l,r}(\omega)-1/2 \end{array}\right) \, .
\end{align}
Here we made use of time translation invariance and assumed a constant density of states in the left and right reservoir
$\rho_{l} = \rho_{r} = \rho_{0}$. The applied finite bias $\mu_{r} - \mu_{l} = eV$ is included in the Fermi distribution functions
$n_{l} = n(\omega - eV/2) = \left[ \exp(\beta (\omega - eV/2) + 1\right]^{-1}$ and $n_{r} = n(\omega + eV/2) = \left[ \exp(\beta (\omega + eV/2) + 1\right]^{-1}$.
The inverse temperature of electrons in the reservoirs is $\beta = 1/k_{B} T$ and we use units where $\hbar = 1$.
%

\subsection{The oscillator}
\label{sec:oscGF}

Since the oscillator modulates the tunneling of electrons and therefore has impact on the measured average current and current-current
correlator, it is important to understand the significance of the oscillator's state. We distinguish between an oscillator in a
stationary state and one in a non-stationary state. We justify this differentiation by arguing that for short times after the
measurement, the oscillator will certainly be non-stationary. The dominating timescale here is the one given by the oscillator
itself, $1/\Omega$, which has to be compared to times scales on which the damping of the oscillator due to the tunnel junction
and the external heat bath happens. In the non-stationary case, we cannot make use of time translation invariance in the oscillator's
correlation function $D(t,t')$. For longer times however, the assumption of stationarity is justified since the oscillator can
equilibrate with the environment and reach a steady state. The oscillator's correlation function now only depends on the
time difference $t-t'$.

We work in the following with the oscillator operators given in the Heisenberg picture as $\hat{x}(t) = \hat{x} \cos(\Omega t) + \hat{p}/(m \Omega) \sin(\Omega t)$
and $\hat{p}(t) = \hat{p} \cos(\Omega t) - \hat{x} (m \Omega) \sin(\Omega t)$. We also define the aforementioned oscillator
correlation function $D(t,t')$ in Keldysh space as
\begin{equation}\label{eqn:d1}
	D(t,t') = -i \ex{T_{c} \, \hat{x}(t) \hat{x}(t')} \, .
\end{equation}
When we later investigate the second order noise we consider both the stationary situation and the non-stationary situation.
The following relation then is a very useful one
\begin{equation}\label{eqn:d2}
	\hat{x}(t+t') = \hat{x}(t')\cos(\Omega t) + \frac{\hat{p}(t')}{m\Omega} \sin(\Omega t) \, .
\end{equation}
For calculations up to second order, we look at the influence of stationary and non-stationary oscillator states
on the current noise, in fourth order we restrict ourselves to the stationary case. Hence, we are interested
in a clear definition of the oscillator's correlation functions and spectral functions in the stationary case, which will be
addressed now.

\subsubsection{Oscillator correlation functions in the stationary case}
\label{sec:oscStat1}

Considering the stationary case, we give useful expressions for the oscillator's correlation functions which later allow us to
identify the oscillator's power spectrum in $x$ denoted by $S_{x}(\omega)$ and in $p$ denoted by $S_{p}(\omega)$. From
Eq.(\ref{eqn:d1}) the correlation function where $t \in C_{+}$ and $t' \in C_{-}$ is given by
\begin{align}\label{eqn:e1}
	&i D^{+-}(t,t') 	=\ex{\hat{x}(t) \hat{x}(t')} = \nn \\
				&=\frac{1}{2} \Big< \bar{x}^{2}_{+} \cos(\Omega (t-t')) + \frac{[\hat{p},\hat{x}]}{m\Omega} \sin(\Omega (t-t')) + \nn \\
				&+\bar{x}^{2}_{-} \cos(\Omega (t+t')) + \frac{\{\hat{p},\hat{x}\}}{m\Omega} \sin(\Omega (t+t')) \Big> \, ,
\end{align}
and similar for $t \in C_{-}$ and $t' \in C_{+}$
\begin{align}\label{eqn:e2}
	&i D^{-+}(t,t') 	=\ex{\hat{x}(t') \hat{x}(t)} = \nn \\
				&=\frac{1}{2} \Big< \bar{x}^{2}_{+} \cos(\Omega (t-t')) - \frac{[\hat{p},\hat{x}]}{m\Omega} \sin(\Omega (t-t')) + \nn \\
				&+\bar{x}^{2}_{-} \cos(\Omega (t+t')) + \frac{\{\hat{p},\hat{x}\}}{m\Omega} \sin(\Omega (t+t')) \Big> \, ,
\end{align}
where we defined
\begin{equation}\label{eqn:e3}
	\bar{x}^{2}_{\pm} = \hat{x}^{2} \pm \frac{\hat{p}^{2}}{m^{2}\Omega^{2}} \, ,
\end{equation}
and $[\cdot,\cdot]$ denotes the commutator and $\{\cdot,\cdot\}$ the anti-commutator. Since, here we deal with the stationary
case, the expectation values $\ex{\bar{x}^{2}_{-}}$ and $\ex{ \{ p,x \} }$ appearing as prefactors of functions depending on $t+t'$
equal zero which one can easily check by using any stationary state, e.g. number-states. As one would expect, the correlation
function now is a function of the time difference $t-t'$ only. The Fourier transform of the correlation functions then yields
\begin{align}
	i D^{+-}(\omega) &= \frac{1}{2} \Big< \bar{x}^{2}_{+} \fr^{+}_{\gamma}(\omega,\Omega) + \frac{i \, [\hat{p}, \hat{x}]}{m\Omega} \fr^{-}_{\gamma}(\omega,\Omega) \Big> \label{eqn:e4} \, , \\
	i D^{-+}(\omega) &= \frac{1}{2} \Big< \bar{x}^{2}_{+} \fr^{+}_{\gamma}(\omega,\Omega) - \frac{i \, [\hat{p}, \hat{x}]}{m\Omega} \fr^{-}_{\gamma}(\omega,\Omega) \Big> \label{eqn:e5} \, .
\end{align}
Additionally, we introduce the two momentum correlation functions $iP^{+-}(t,t') = \ex{\hat{p}(t) \hat{p}(t')}$ and
$iP^{-+}(t,t') = \ex{ \hat{p}(t') \hat{p}(t)}$. The same arguments as for $i D^{\pm \mp}(t,t')$ lead here to the following Fourier
transforms
\begin{align}
	i P^{+-}(\omega) &= \frac{1}{2} \Big< \bar{p}^{2}_{+} \fr^{+}_{\gamma}(\omega,\Omega) + m \Omega i \, [\hat{p}, \hat{x}] \fr^{-}_{\gamma}(\omega,\Omega) \Big> \label{eqn:e6} \, , \\
	i P^{-+}(\omega) &= \frac{1}{2} \Big< \bar{p}^{2}_{+} \fr^{+}_{\gamma}(\omega,\Omega) - m \Omega i \, [\hat{p}, \hat{x}] \fr^{-}_{\gamma}(\omega,\Omega) \Big> \label{eqn:e7} \, ,
\end{align}
where similar to above
\begin{equation}\label{eqn:e8}
	\bar{p}^{2}_{\pm} = m^{2}\Omega^{2} \hat{x}^{2} \pm \hat{p}^{2} \, .
\end{equation}
We introduce the functions $\fr^{\pm}_{\gamma}(\omega,\Omega)$ as
\begin{align}
	\fr^{+}_{\gamma \rightarrow 0}(\omega,\Omega) &= \pi \left[ \delta(\omega+\Omega) + \delta(\omega-\Omega) \right] \label{eqn:e9a} \, , \\
	\fr^{-}_{\gamma \rightarrow 0}(\omega,\Omega) &= \pi \left[ \delta(\omega-\Omega) - \delta(\omega+\Omega) \right] \label{eqn:e9b} \, .
\end{align}
The coupling of the oscillator to two environments, namely an external heat bath and the tunnel junction, being at the
temperatures $T_{\rm env}$ and $k_{B} T_{\rm junc} = eV/2$ \cite{Mozyrsky:2002p135} respectively, introduces a damping of
the oscillator with damping coefficients  $\gamma_{0}$ and $\gamma_{+}$ respectively.
The oscillator dynamics due to the coupling to the tunnel junction can be calculated by solving a Dyson equation for
the oscillator correlation function $D(t,t')$ where the self-energy is taken to lowest non-vanishing order in the tunnel Hamiltonian,
i.e. $\Sigma(t,t') = -i \ex{T_{c} \, \hat{h}_{1}(t) \hat{h}_{1}(t')}$. Using the Keldysh technique as done in Refs.~\onlinecite{Wabnig:2007p16,Clerk:2004p154}
the oscillator dynamics and the damping coefficient $\gamma_{+} = \pi \rho_{0}^{2} t_{1}^{2} / m$ can be calculated.
The coupling to the external heat bath can be added phenomenologically, or particularly as an interaction with a bath of
harmonic oscillators. The total damping then follows as $\gamma_{\rm tot} = \gamma_{0} + \gamma_{+}$. We can assign
an effective temperature $T_{\rm eff}$ to the oscillator with $\gamma_{\rm tot} \, T_{\rm eff} = \gamma_{+} \, T_{\rm junc} + \gamma_{0} \, T_{\rm env}$.
This leads to the general case for $\fr^{\pm}_{\gamma}$ by replacing the $\delta$-functions in Eqs.~(\ref{eqn:e9a},\ref{eqn:e9b}) by a Lorentzian, where we
include both sources of damping and an oscillator frequency $\Omega \rightarrow \sqrt{\Omega^{2}-\gamma^{2}}$.
For this damped case we can write for $\fr^{+}_{\gamma}(\omega,\Omega)$ and $\fr^{-}_{\gamma}(\omega,\Omega)$
\begin{align}
	\fr^{+}_{\gamma}(\omega,\Omega) &= \frac{2 \gamma_{\rm tot} (\omega^{2} + \Omega^{2})}{4 \gamma_{\rm tot}^{2} \omega^{2} +(\omega^{2} - \Omega^{2})^{2}} \label{eqn:e10} \, , \\
	\fr^{-}_{\gamma}(\omega,\Omega) &= \frac{4 \gamma_{\rm tot} \, \omega \, \sqrt{ \Omega^{2} - \gamma_{\rm tot}^{2}}}{4 \gamma_{\rm tot}^{2} \omega^{2} +(\omega^{2} - \Omega^{2})^{2}} \label{eqn:e10b} \, .
\end{align}
We want to stress that we have not made any assumption on the initial time as e.g. $t'=0$. With the above made definitions,
the following relation (which has to hold for any bosonic correlation function in Keldysh space)
\begin{equation}\label{eqn:e11}
	D^{ij}(\omega) = D^{ji}(-\omega)
\end{equation}
can easily be verified, here $i$ and $j$ are Keldysh indices. This concludes our discussion on the oscillator correlation
functions. We now turn to the spectral functions $S_{x}(\omega)$ and $S_{p}(\omega)$.

\subsubsection{The oscillator's spectra in the stationary case}
\label{sec:oscStat2}

The symmetrized power spectrum, in general defined as $\frac{1}{2} \int \, dt \, e^{i\omega t} \ex{ \{ \hat{\varUpsilon}(t), \hat{\varUpsilon}(t') \} }$
of the oscillator quantities $\hat{\varUpsilon} = \hat{x}, \hat{p}$ is an observable that can be measured by e.g. current noise measurements
(as discussed below). Both $S_{x}(\omega)$ and $S_{p}(\omega)$ can be measured through the current noise $S_{\rm sym}(\omega)$. The
expressions for these power spectra are given by
\begin{align}\label{eqn:f1}
	S_{x}(\omega) 	&= \frac{1}{2} \int \, dt \, e^{i\omega t} \ex{ \{ \hat{x}(t), \hat{x}(t') \} } = \nn \\
				&= \frac{1}{2} \int \, dt \, e^{i\omega t} \, i \, \left[ D^{+-}(t,t') + D^{-+}(t,t') \right] = \nn \\
				&= \frac{1}{2} i \, \left[ D^{+-}(\omega) + D^{-+}(\omega) \right] = \nn \\
				&= \frac{1}{2} \ex{ \bar{x}^{2} } \fr^{+}_{\gamma}(\omega,\Omega) \, ,
\end{align}
and
\begin{align}\label{eqn:f2}
	S_{p}(\omega) 	&= \frac{1}{2} \int \, dt \, e^{i\omega t} \ex{ \{ \hat{p}(t), \hat{p}(t') \} } = \nn \\
				&= \frac{1}{2} \int \, dt \, e^{i\omega t} \, i \, \left[ P^{+-}(t,t') + P^{-+}(t,t') \right] = \nn \\
				&= \frac{1}{2} i \, \left[ P^{+-}(\omega) + P^{-+}(\omega) \right] = \nn \\
				&= \frac{1}{2} \ex{ \bar{p}^{2} } \fr^{+}_{\gamma}(\omega,\Omega) \, .
\end{align}
The momentum and position spectrum are related via the relation
\begin{align}\label{eqn:f3}
	S_{p}(\omega) = m^{2} \Omega^{2} S_{x}(\omega) \, .
\end{align}
We can also write down the spectra using the Keldysh Green's function $D^{K}(\omega)=D^{+-}(\omega)+D^{-+}(\omega)$
which yields
\begin{align}\label{eqn:f4}
	S_{x}(\omega) &= \frac{1}{2} \, i D^{K}(\omega) \, ,\\
	S_{p}(\omega) &= \frac{1}{2} m^{2} \Omega^{2} \, i D^{K}(\omega) \, .
\end{align}
To further simplify the notation, we introduce
\begin{align}\label{eqn:f5}
	Q(\omega) 	&= \frac{i}{2} \left[ D^{+-}(\omega) - D^{-+}(\omega) \right] = \nn \\
				&= \frac{i}{2} \left[ P^{+-}(\omega) - P^{-+}(\omega) \right] = \nn \\
				&= \frac{i}{2} \frac{\left[ \hat{p}, \hat{x} \right] }{m \Omega} \fr^{-}_{\gamma}(\omega,\Omega) = \nn \\
				&= \frac{1}{2 m \Omega} \fr^{-}_{\gamma}(\omega,\Omega) \, ,
\end{align}
which is used later in the fourth order noise calculation.

\section{Current noise calculations}
\label{sec:perturb}

In this section, we cover a variety of aspects when dealing with the current noise. For all different aspects we find expressions
for the noise which are valid for an arbitrary $\eta$ and therefore include the $x$-detector as well as the $p$-detector.

The first part is dedicated to the noise in second order perturbation theory, where we furthermore make the distinction
between a stationary harmonic oscillator and a non-stationary one. Beside the Markovian regime ($eV \gg \hbar \Omega$), the
Keldysh formalism also allows us to investigate the non-Markovian regime ($eV \ll \hbar \Omega$). The main results in this
section are that the current noise for a non-stationary oscillator can in principle be complex. In this case, a detectable complex
noise would allow for a nearly complete determination of the oscillators covariance matrix $\sigma_{ij} = \tr{(\hat{\rho} \, \{ \hat{\varUpsilon}_{i},\hat{\varUpsilon}_{j}\}/2)}$,
where $\hat{\varUpsilon} = (\hat{x},\hat{p})^{T}$. The covariance matrix allows for a complete description of the oscillator's
quantum state.

For the stationary oscillator we recover a noise that is real and in accordance with the Wiener-Khinchin theorem. This noise
is the well know noise of a bare biased tunnel junction which shows kinks at $|\omega| = |V|$ \cite{Blanter:2000p98,Schoelkopf:1997p306},
modified by the oscillator leading to kinks at $|\omega| = |V \pm \Omega|$.

In the last part of this section, we deal with the noise up to fourth order in the tunneling amplitudes. Then we restrict ourselves
on the stationary case. On the one hand, the fourth order contributions modify the kinks, on the other hand, they give rise to
resonances stemming from the oscillator correlation functions \cite{Clerk:2004p185,Wabnig:2007p16,Doiron:2008p13}.

We now introduce the perturbation theory leading to the noise expression.

\subsection{Overview}
\label{sec:}

In general, the expression for the current-current correlator in the Keldysh formalism is given by
\begin{align}\label{eqn:g1}
	S(\tau_{3},\tau_{4}) &= \ex{ T_{c} \, e^{-i \int_{c} d\tilde{\tau} \hat{H}_{\rm tun}(\tilde{\tau})} \hat{I}(\tau_{3}) \hat{I}(\tau_{4}) } - \nn \\
	&-\ex{\hat{I}(\tau_{3})} \ex{\hat{I}(\tau_{4})} \, ,
\end{align}
since we consider only the second and fourth order current noise we write
\begin{equation}\label{eqn:g2}
	S(\tau_{3},\tau_{4}) = S^{(2)}(\tau_{3},\tau_{4}) + S^{(4)}(\tau_{3},\tau_{4}) \, ,
\end{equation}
where
\begin{equation}\label{eqn:g3}
	S^{(2)}(\tau_{3},\tau_{4}) = \ex{T_{c} \, \hat{I}(\tau_{3}) \hat{I}(\tau_{4}) } \, ,
\end{equation}
and
\begin{align}\label{eqn:g4}
	&S^{(4)}(\tau_{3},\tau_{4}) = \nn \\
		&=-\frac{1}{2} \int_{c} d\tau_{1} d\tau_{2} \ex{ T_{c} \, \hat{H}_{\rm tun}(\tau_{1}) \hat{H}_{\rm tun}(\tau_{2}) \hat{I}(\tau_{3}) \hat{I}(\tau_{4}) } + \nn \\
		&+ \int_{c} d\tau_{1} d\tau_{2} \ex{ T_{c} \, \hat{H}_{\rm tun}(\tau_{1})\hat{I}(\tau_{3}) } \ex{ T_{c}\hat{H}_{\rm tun}(\tau_{2})\hat{I}(\tau_{4}) } \, .
 \end{align}
The general expression for the average current is given by
\begin{equation}\label{eqn:g5}
	\ex{I(t)} = \ex{T_{c} \, e^{-i\int_{c} d\tau \hat{H}_{\rm tun}(\tau)} \hat{I}(t)} \, .
\end{equation}
To second order in the tunneling amplitudes, the average current can be calculated by
\begin{equation}\label{eqn:g6}
	\ex{I(t)} = -i \int_{c} d\tau \ex{T_{c} \, \hat{H}_{\rm tun}(\tau) \hat{I}(t)} \, ,
\end{equation}
and we obtain
\begin{align}\label{eqn:g6a}
	\ex{I(t)} = & 2 \pi \rho_{0}^{2} e \Big\{ t_{0}^{2} e V + \nn \\
			+& 2 \cos(\eta) t_{0} t_{1} e V \ex{\hat{x}(t)} + \sin(\eta) t_{0} t_{1} \frac{\ex{\hat{p}(t)}}{m} + \nn \\
			+& t_{1}^{2} e V \ex{\hat{x}(t)\hat{x}(t)} - \frac{t_{1}^{2}}{2 m \Omega} \sigma^{-}(\Omega,V) \Big\} \, ,
\end{align}
where $\sigma^{-}(\Omega,V)$ is given in Eq.~(\ref{eqn:i2}). Our result for the average current is in accordance with Ref.~\onlinecite{Doiron:2008p13}.
The current noise we calculate is always the frequency-dependent (and in the non-stationary case also time-dependent) symmetrized
current noise, defined as
\begin{equation}\label{eqn:g7}
	S_{\rm sym}(\omega,t') = \frac{1}{2} \int dt \, e^{i \omega t} \left[ S^{-+}(t,t') + S^{+-}(t,t') \right]
\end{equation}
where for $S^{-+}(t,t')$, $t \in C_{-}$ and $t' \in C_{+}$ and similar for $S^{+-}(t,t')$, here $C_{-}$ and $C_{+}$ are the lower
and upper branch of the Keldysh contour $C$, respectively, see Fig.~\ref{fig:keldyshC}.

\subsection{Current noise to second order in the tunneling amplitudes}

We now turn to the calculation of the current noise to second order. With the current operator $\hat{I}$ already being first order in
the tunneling amplitudes, the current noise in the Born approximation is given by the following expression
\begin{align}\label{eqn:h1}
	S^{(2) ij}(t,t') = \ex{ T_{c} \, \hat{I}(t+t')_{i} \hat{I}(t')_{j} } \, ,
\end{align}
where we used a slightly different definition of the current noise. This definition will be useful when examining the non-stationary
case. Due to this definition the time dependance on $t'$ in the symmetrized current noise is only present in the oscillator's
quantum mechanical expectation values.

The general expression for the current noise in Keldysh space reads
\begin{align}\label{eqn:h2}
	S^{(2) ij}(t,t') &= e^{2} \big[ \fg_{00}(t+t',t') + \nn \\
			&+ \ex{\hat{x}(t')} \fg_{01}(t+t',t') + \nn \\
			&+ \ex{\hat{x}(t+t')} \fg_{10}(t+t',t') + \nn \\
			&+ i D(t+t',t') \, \fg_{11}(t+t',t')\big] \, .
\end{align}
where $\fg_{ij}(t,t')$ is given in App.~\ref{sec:appendix1}.

\subsubsection{General expression for the current noise}

In this section, we only make use of time translation invariance in the $\fg_{ij}(t,t')$ functions, the oscillator is taken as
non-stationary. Details of the calculation can be found in App.~\ref{sec:appendix2}. The final result we obtain for the
symmetrized current noise to second order in the tunneling amplitudes reads
\begin{widetext}
\begin{align}\label{eqn:i1}
S^{(2)}_{\rm sym}(\w,t')	&= 2 \pi \rho_{0}^{2} e^{2} \Big\{ t_{0}^{2} \sigma^{+}(\omega,V) + \nn \\
				&+ \ex{\hat{x}(t')}  t_{0}t_{1} \cos(\eta) \left[  \sigma^{+}(\omega,V)+ \frac{1}{2} \left( \sigma^{+}(\omega +\Omega,V) + \sigma^{+}(\omega-\Omega,V)\right) \right] - \nn \\
				&- \ex{\hat{x}(t')} t_{0}t_{1} i \sin(\eta) \left[ \sigma^{-}(\omega,V) - \frac{1}{2} \left( \sigma^{-}(\omega +\Omega,V) + \sigma^{-}(\omega-\Omega,V) \right) \right] - \nn \\
				&- \ex{\hat{p}(t')} \frac{t_{0} t_{1}}{2m\Omega} \Big[ \sin(\eta) \left[ \sigma^{-}(\omega +\Omega,V) - \sigma^{-}(\omega-\Omega,V) \right]  - i \cos(\eta) \left[ \sigma^{+}(\omega +\Omega,V) - \sigma^{+}(\omega-\Omega,V) \right] \Big]  + \nn \\	
				&+ \ex{\hat{x}(t') \hat{x}(t')} \frac{t_{1}^{2}}{2} \Big[ \sigma^{+}(\omega +\Omega,V) + \sigma^{+}(\omega-\Omega,V) \Big] - \frac{t_{1}^{2}}{2 m} - \nn \\
				&- \ex{ \{ \hat{x}(t'),\hat{p}(t') \} } \frac{i \, t_{1}^{2}}{4 m \Omega} \Big[ \sigma^{+}(\omega +\Omega,V) - \sigma^{+}(\omega-\Omega,V) \Big] \Big\}  \, ,
\end{align}
\end{widetext}
where we separated the real and imaginary part using the relative phase $\eta$ between the tunneling amplitudes
and in addition introduced
\begin{align}\label{eqn:i2}
	\sigma^{\pm}(\xi,V) = & \frac{eV+\xi}{2} \coth\left(\beta \frac{eV+\xi}{2}\right) \nn \\ &\pm \frac{eV-\xi}{2} \coth\left(\beta \frac{eV-\xi}{2}\right) \, .
\end{align}
We want to note the important aspect of the current noise that it possibly can have a complex valued character which we discuss later.
The gained expression is quite lengthly but provides us with the full quantum mechanical non-equilibrium characteristics
of the current noise in the Markovian as well as in the non-Markovian regime. We made no assumptions on the state of the oscillator,
which now gives us the possibility to identify momentum properties of the nanomechanical resonator using the current noise spectrum
$S^{(2)}_{\rm sym}(\omega,t')$. In the next section, we discuss this new possibility of a $p$-detector which involves measuring
a complex valued current noise.

\subsubsection{Complex current noise and the p-detector in the non-stationary case}
\label{sec:meas}

The expression in Eq.~(\ref{eqn:i1}) allows for a comparison with results obtained in Ref.~\onlinecite{Doiron:2008p13} where
it was possible with a phase of $\eta = \pi/2$ to determine the momentum of the oscillator. In Ref.~\onlinecite{Doiron:2008p13}
an Aharonov-Bohm setup allows the tuning of the relative phase $\eta$. The full current noise spectrum there is proportional
to the position spectrum $S_{x}(\omega)$ which is peaked at $\omega = \pm \Omega$ in the case of $\eta=0$ and in the case
of $\eta=\pi/2$ is proportional to the momentum spectrum $S_{p}(\omega)$ showing peaks at $\omega = \pm \Omega$. This
peaked structure of the current noise spectrum is a fourth order effect, as we will see and discuss later when dealing
with the fourth order corrections to the current noise.

As one can see from Eq.~(\ref{eqn:i1}), already the second order current noise allows to determine the expectation value of the
oscillator's momentum and in addition to that of the anticommutator $\{ \hat{x},\hat{p} \}$, even if the phase $\eta = 0$, i.e. the
Aharonov-Bohm setup becomes obsolete in our case. The signature of the oscillator's momentum $\hat{p}$ in our case is
however different than the one in Ref.~\onlinecite{Doiron:2008p13}. Instead of the peaked structure, we find a kink-like structure
which stems from the fact that we deal with second order perturbation theory.

In order to understand how one can use this to identify the momentum, we have to understand the meaning of a complex current noise.
As stated in Ref.~\onlinecite{DiCarlo:2009p128} a complex valued current noise is in principle a measurable quantity. To have a relevant
measurable quantity we would have to average the time dependent current noise $S^{(2)}_{\rm sym}(\omega,t')$ over the measurement
time $\Delta T$. We could do this in the following way
\begin{equation}\label{eqn:j1}
	\bar{S}^{(2)}_{\rm sym}(\omega) = \frac{1}{\Delta T} \int_{-\Delta T/2}^{\Delta T/2} dt' S^{(2)}_{\rm sym}(\omega,t') \, .
\end{equation}
Since the time dependance of the current noise is only visible in the expectation values of the oscillator's variables, it is important for
the actual measurement to consider the time scales which are involved. If the measurement time $\Delta T$ is less than the
time scale of the oscillator ($1/\Omega$), the measured time averaged current noise $\bar{S}_{\rm sym}^{(2)}(\omega)$ will be time-dependent.
If however, the oscillator undergoes multiple oscillation cycles during the measurement time, the current noise will be time-independent.
In this case we could as well take the oscillator to be in a stationary state. For a damped oscillator the times scales on which the
damping happens have to be taken into account as mentioned already in Sec.~\ref{sec:oscGF}.

We conclude with remarks on the interesting non-stationary case, where we can also take $\eta = 0$ without losing the
information on $\ex{\hat{p}(t')}$ and in addition obtain information on $\ex{\{ \hat{x}(t'),\hat{p}(t') \}}$. We intend to give an idea
of how to access the information on $\ex{\hat{p}(t')}$ and $\ex{\{ \hat{x}(t'),\hat{p}(t') \}}$ available through the complex current noise.

The expectation value of $\ex{\hat{p}(t')}$ with respect to number states or a linear combination of them, will always vanish
when averaging over time according to Eq.~(\ref{eqn:j1}). However, this is different for coherent states
$\ket{\alpha} = \exp(-|\alpha|^{2}/2) \, \sum_{n=0}^{\infty} \alpha^{n}/n! \ket{n}$, where we can write $\alpha = |\alpha| \exp(i \, \delta)$
with $|\alpha|$ being the amplitude and $\delta$ the phase of the coherent state, respectively. The time averaged expectation
values $\ex{\hat{p}(t')}$ and $\ex{\{ \hat{x}(t'),\hat{p}(t') \}}$ with respect to $\ket{\alpha}$ yield
\begin{align}\label{eqn:j2}
	\ex{\hat{p}(t')}_{av} &= \sqrt{2 m \Omega } \frac{2 |\alpha|}{\Omega \Delta T} \sin(\delta) \sin(\Omega \Delta T /2) \, ,
\end{align}
and
\begin{align}\label{eqn:j3}
	\ex{\{ \hat{x}(t'),\hat{p}(t') \}}_{av} &= \frac{2 |\alpha|^{2}}{\Omega \Delta T} \sin(2\delta) \sin(\Omega \Delta T) \, .
\end{align}
For short measurement times $\Delta T < 1/\Omega$ we can write
\begin{align}\label{eqn:j4}
	\lim_{\Delta T \rightarrow 0} \ex{\hat{p}(t')}_{av} = \sqrt{2 m \Omega} |\alpha| \sin(\delta) \, ,
\end{align}
\begin{align}\label{eqn:j5}
	\lim_{\Delta T \rightarrow 0} \ex{\{ \hat{x}(t'),\hat{p}(t') \}}_{av}  = 2 |\alpha|^{2} \sin(2 \delta) \, .
\end{align}
We sparate the current noise $\bar{S}^{(2)}_{\rm sym}(\omega) = \bar{S}^{(2)}_{\rm sym,R}(\omega) + \bar{S}^{(2)}_{\rm sym,I}(\omega)$ into
real and imaginary part where we observe that the imaginary part $\bar{S}^{(2)}_{\rm sym,I}(\omega)$ only contains information
on the oscillator's momentum $\ex{\hat{p}(t')}$ and the anticommutator $\ex{\{ \hat{x}(t'),\hat{p}(t') \}}$
\begin{align}\label{eqn:j6}
	& \bar{S}^{(2)}_{\rm sym,I}(\omega) = 2 \pi \rho_{0}^{2} e^{2} \Big\{ \nn \\
							& \frac{1}{\sqrt{ 2 m \Omega}} t_{0} t_{1} |\alpha| \sin(\delta) \left[ \sigma^{+}(\omega +\Omega,V) - \sigma^{+}(\omega-\Omega,V) \right] - \nn \\
							&- \frac{t_{1}^{2} |\alpha|^{2}}{2 m \Omega} \sin(2 \delta) \left[ \sigma^{+}(\omega +\Omega,V) - \sigma^{+}(\omega-\Omega,V) \right] \Big\} \, .
\end{align}
The phase $\delta$ of the coherent state now allows for a determination of the oscillator's momentum. For $\delta=\pi/2$,
the signature in the imaginary part of the time averaged noise $\bar{S}^{(2)}_{\rm sym,I}(\omega)$ stems only from the oscillator's
momentum. The signal in the non-Markovian regime is more pronounced than in the Markovian regime, see Eq.~(\ref{eqn:j6}).

\subsubsection{Current noise in the stationary case}

Contrary to the non-stationary case we now also assume time translation invariance in the oscillator correlation function, i.e.
$D(t,t') = D(t-t')$. One can see that the calculation in the stationary case goes along the same lines as in the non-stationary
case. The only difference will be that oscillator expectation values are now taken at time $t'=0$, i.e. we encounter for instance
$\ex{\hat{x}(0)}$ instead of $\ex{\hat{x}(t')}$.

When interpreting the result for the stationary case we have to keep the constrains on oscillator expectation values in mind.
These constrains mentioned in Sec.~\ref{sec:oscStat1} lead to vanishing expectation values of the anticommutator
$\ex{\{ \hat{x}, \hat{p} \}}$ and vanishing expectation values for $\ex{\hat{x}}$ and $\ex{\hat{p}}$. The current noise to second order
is then equivalent to the ones previously obtained in Refs.~\onlinecite{Wabnig:2007p16, Schmidt:2010p301},
cf. Eq.~(C.4) in Ref.~\onlinecite{Schmidt:2010p301} with $\gamma_{2} = \ex{x} = \ex{p} =0$.

\subsection{Current noise to fourth order in the tunneling amplitudes}

\begin{figure}[ht]
	\center{\includegraphics[width=1\columnwidth]{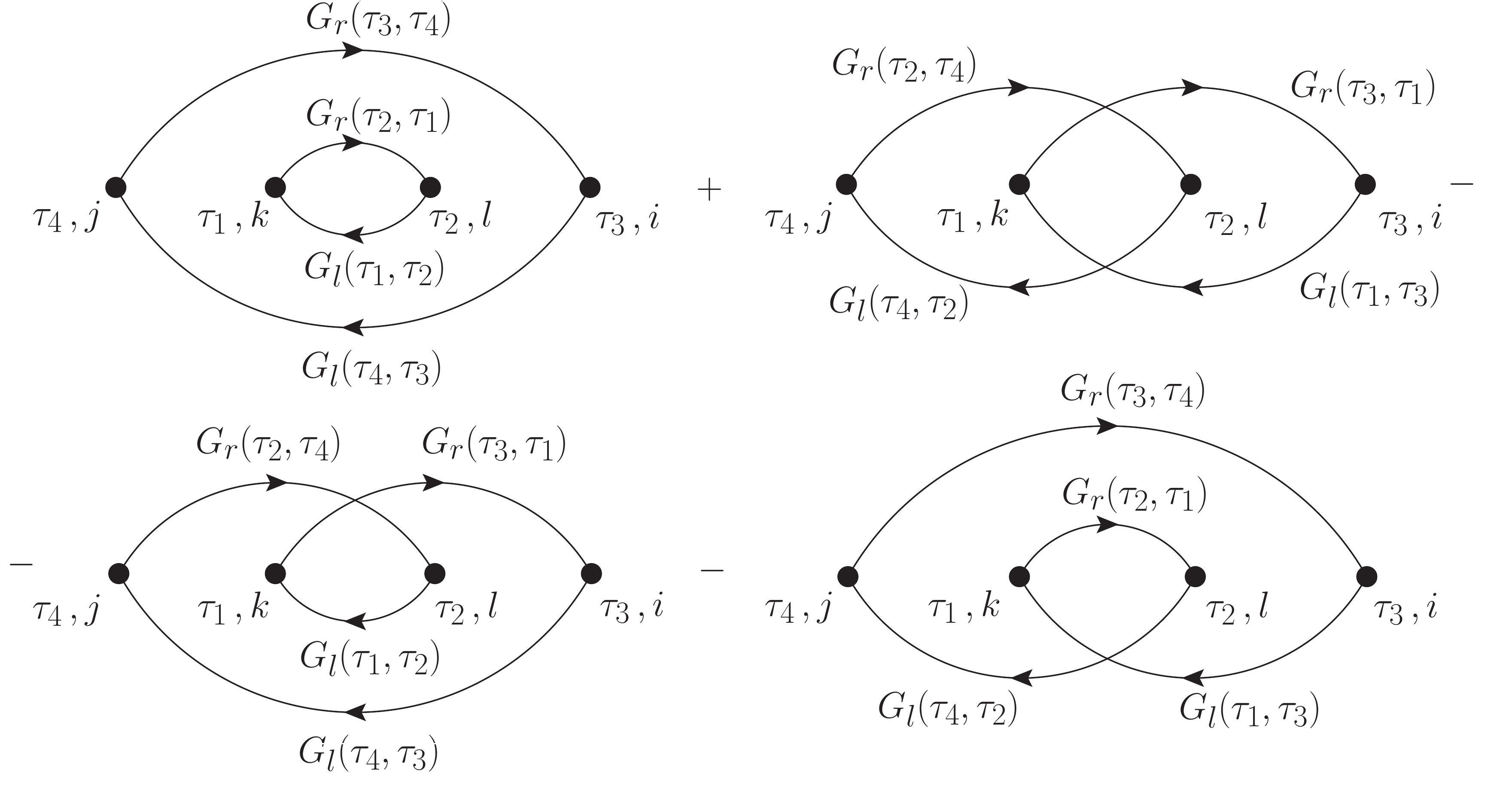}}
	\caption{Diagramatic representation for Eq.~(\ref{eqn:k2}) where we omitted the factor $\beta^{*}_{i_{1}} \beta_{i_{2}} \beta_{i_{3}} \beta^{*}_{i_{4}}$, note that
			Eq.~(\ref{eqn:k2}) only contains fermionic Green's functions. }
	\label{fig:diadet}
\end{figure}

\begin{figure}[ht]
	\center{\includegraphics[width=\columnwidth]{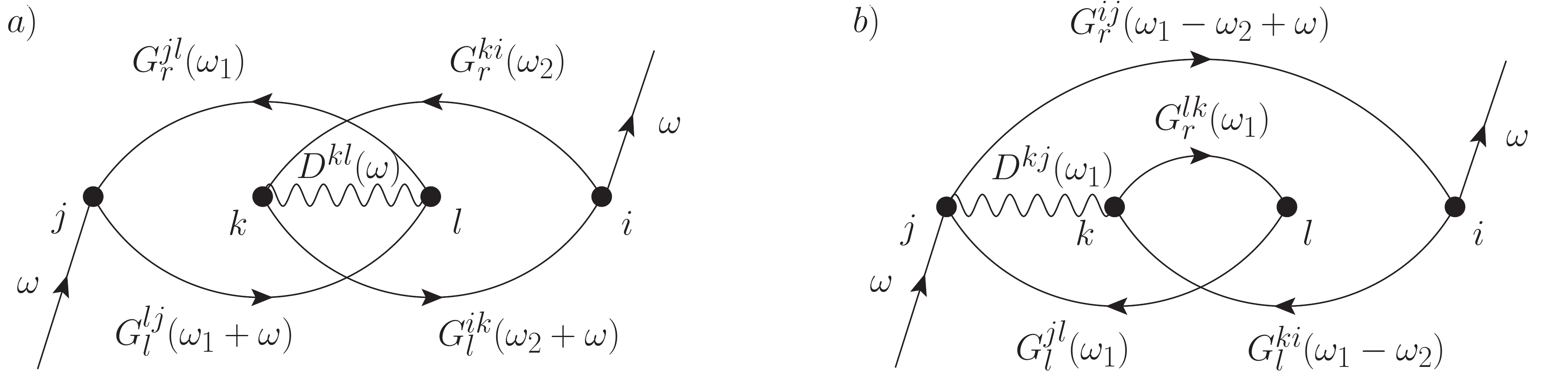}}
	\caption{ Examples of diagrams containing one oscillator correlation function $D$. In panel $a)$ a bubble diagram is depicted where the
			oscillator line is independent of an internal frequency. Panel $b)$ shows a closed loop diagram where the oscillator line always
			appears under integration of an internal frequency.  }
	\label{fig:diaosc}
\end{figure}

We now turn to the investigation of the fourth order current noise. Since the fourth order perturbation theory involves a large
amount of terms we use a diagramatic approach. In the case of the fourth order current noise we restrict ourself to the stationary case
for simplicity. An overview of all contributing terms in the non-stationary case is given in App.~\ref{sec:appendix3}. In what
follows we give a short explanation of the diagramatics. From Eq.~(\ref{eqn:g4}) it becomes obvious that $S^{(4)}$ contains
fermionic expectation values which have the form
\begin{align}\label{eqn:k1}
	& \fm_{\rm i_{1},i_{2},i_{3},i_{4}}(\tau_{1},\tau_{2},\tau_{3},\tau_{4}) = \nn \\
	&= \ex{ T_{c} \, \hat{h}_{i_{1}}(\tau_{1}) \hat{h}_{i_{2}}(\tau_{2}) \hat{j}_{i_{3}}(\tau_{3}) \hat{j}_{i_{4}}(\tau_{4}) } = \nn \\
	&= \Big< T_{c} \, \ft^{\dag}_{i_{1}}(\tau_{1}) \ft_{i_{2}}(\tau_{2}) \ft_{i_{3}}(\tau_{3}) \ft^{\dag}_{i_{4}}(\tau_{4}) + \nn \\
	&+ (3\leftrightarrow 4) - (2\leftrightarrow 4) + h.c. \Big> \, .
\end{align}
The index $i_{j}$ determines whether we are dealing with $\beta_{0}$ or $\beta_{1}$, $i_{j} \in \{0,1\}$. It is only necessary to
evaluate the first expectation value in Eq.~(\ref{eqn:k1}) using Wick's theorem, the other ones follow as indicated by
$(3\leftrightarrow 4)$, $(2\leftrightarrow 4)$ and hermitian conjugation. This first term leads to
\begin{align}\label{eqn:k2}
	&\ex{ T_{c} \, \ft^{\dag}_{i_{1}}(\tau_{1}) \ft_{i_{2}}(\tau_{2}) \ft_{i_{3}}(\tau_{3}) \ft^{\dag}_{i_{4}}(\tau_{4}) } = \nn \\
	&= \beta^{*}_{i_{1}} \beta_{i_{2}} \beta_{i_{3}} \beta^{*}_{i_{4}} \times \nn \\
	& \times \left[ G_{l}(\tau_{1},\tau_{2}) G_{l}(\tau_{4},\tau_{3}) - G_{l}(\tau_{1},\tau_{3}) G_{l}(\tau_{4},\tau_{2}) \right] \times \nn \\
	& \times \left[ G_{r}(\tau_{2},\tau_{1}) G_{r}(\tau_{3},\tau_{4}) - G_{r}(\tau_{3},\tau_{1}) G_{r}(\tau_{2},\tau_{4}) \right] \, .
\end{align}
Similar to Ref.~\onlinecite{Wabnig:2007p16} we use a diagrammatic representation for the expression in Eq.~(\ref{eqn:k1}).
As an example,we show the diagrams emerging from the expression in Eq.~(\ref{eqn:k2}) in Fig.~\ref{fig:diadet} and explain
the components of the diagrams.
Fermionic Green's functions of the reservoirs are represented by solid lines and vertices are depicted by a dot, labeled with
a time variable and a Keldysh index indicating on which branch of the Keldysh contour the time lies. An integration over
internal times $\tau_{1}$ and $\tau_{2}$ is implicit. In addition we also have to sum over the two internal Keldysh indices $k$ and $l$.
We recognize that we have to deal with two types of diagrams: diagrams which consist of one closed fermion loop (diagrams
in the lower panel of Fig.~\ref{fig:diadet}) and diagrams consisting of two closed fermion loops/bubbles (diagrams in the
upper panel of Fig.~\ref{fig:diadet}). These two different types of diagrams give very different contributions to the current noise
which we will discuss below. We include the oscillator correlation function $D(t,t')$ in diagrams by a wiggly line connecting two
vertices. In Fig.~\ref{fig:diaosc}, we give an example of diagrams in frequency space containing one oscillator correlation function.
Here, integration over the two internal frequencies $\omega_{1}$ and $\omega_{2}$ as well as summation over the internal
Keldysh indices $k$ and $l$ is implied. In frequency space the difference between the closed loop diagrams, $b)$ in Fig.~\ref{fig:diaosc}
and the bubble diagrams, $a)$ in Fig.~\ref{fig:diaosc} becomes clear: for the closed loop diagrams, the oscillator line always
appears under integration of an internal frequency, whereas for the bubble diagrams, there is no integration over the oscillator line.
This is the reason for the different kind of contribution to the current noise of closed loop and bubble diagrams. As we will show below,
bubble diagrams lead to peaks in the current noise, whereas closed loop diagrams lead to the afore mentioned kinks in the current noise.
We reduce the number of diagrams by only keeping contributions $\sim t_{0}^{2} t_{1}^{2}$ and $\sim t_{1}^{4}$ which are
the only finite contributions in the stationary case. Details are given in App.~\ref{sec:appendix3}. This allows us to
write the current noise $S^{(4)}(\tau_{3},\tau_{4})$ for the further analysis as
\begin{align}\label{eqn:k3}
	S^{(4)}(\tau_{3},\tau_{4}) &= \hat{S}^{(4)}_{D}(\tau_{3},\tau_{4}) + \check{S}^{(4)}_{D}(\tau_{3},\tau_{4}) + \nn \\
		&+ \hat{S}^{(4)}_{DD}(\tau_{3},\tau_{4}) + \check{S}^{(4)}_{DD}(\tau_{3},\tau_{4}) \, ,
\end{align}
where $\hat{S}$ includes the bubble diagrams, $\check{S}$ includes closed loop diagrams and $D$ indicates the number of
oscillator lines in the diagrams. The final result we obtain is valid for an arbitrary relative phase $\eta$ which goes beyond the
result obtained by Wabnig {\it et al.} in Ref.~\onlinecite{Wabnig:2007p16}. This fact allows us to study the $p$-detector in fourth
order perturbation theory.

\subsubsection{ Results for $\hat{S}^{(4)}_{D}(\tau_{3},\tau_{4})$, $\check{S}^{(4)}_{D}(\tau_{3},\tau_{4})$, $\hat{S}^{(4)}_{DD}(\tau_{3},\tau_{4})$ and $\check{S}^{(4)}_{DD}(\tau_{3},\tau_{4})$ }

In the following, we sum up the different types of diagrams, bubble type diagrams as well as closed loop diagrams, both
then can be integrated exactly.

First we consider all diagrams containing only one oscillator line, these diagrams are all proportional to $t_{0}^{2} \, t_{1}^{2}$
and depend on $\eta$. We find for the symmetrized frequency dependent current noise
\begin{align}\label{eqn:l1}
	& \hat{S}^{(4)}_{\rm sym,D}(\omega)	= 4 \pi^{2} e^{2} \rho_{0}^{4} \, t_{0}^{2}t_{1}^{2} \, \Bigg\{ \nn \\
							& \cos(\eta)^{2} \left[ 4 e^{2} V^{2} - 4eV \sigma^{-}(\omega,V) \, \frac{Q(\omega)}{S_{x}(\omega)} \right] \, S_{x}(\omega) + \nn \\
							+& \sin(\eta)^{2} \left[ \omega^{2} - 2\omega \sigma^{+}(\omega,V) \, \frac{Q(\omega)}{S_{x}(\omega)} \right] \, S_{x}(\omega) + \nn \\
							+& \frac{1}{2} \, \left[ D^{R}(\omega) + D^{A}(\omega) \right] \cos(\eta) \sin(\eta) \times \nn \\ & \qquad \times \left[ 4eV \sigma^{+}(\omega,V) - 2\omega \sigma^{-}(\omega,V) \right] \Bigg\} \, ,
\end{align}
which is one of the main results of this paper.

In the case of the closed loop diagrams containing one oscillator line, it is also possible to sum up all diagrams and integrate
them exactly. The expression for $\check{S}^{(4)}_{\rm sym,D}(\omega)$ is rather lengthly, therefore we do not present it here.

We find that the current noise signature of $\check{S}^{(4)}_{\rm sym,D}(\omega)$ is of the kink-like structure similar to
$S^{(2)}_{\rm sym,D}(\omega)$. In addition to the kinks at $|\omega|  = |V \pm \Omega|$ coming from $S^{(2)}_{\rm sym,D}(\omega)$,
$\check{S}^{(4)}_{\rm sym,D}(\omega)$ gives rise to extra kinks at $|\omega| = |V|$ and $|\omega| = |\Omega|$. However, these contributions are only minor modifications
to the current noise floor $S^{(2)}_{\rm sym,D}(\omega)$. Experiments as in Ref.~\onlinecite{FlowersJacobs:2007p188} focus on
the current noise near the resonance frequency $\omega \approx \Omega$ for which $\hat{S}^{(4)}_{\rm sym,D}(\omega)$ is the
most important contribution. Therefore, the discussion of our result will focus on the contributions stemming from
$\hat{S}^{(4)}_{\rm sym,D}(\omega)$. These contributions posses a peaked structure, since the oscillator correlation functions
$D^{R/A}(\omega)$ and the spectrum $S_{x}(\omega)$ is peaked around $\omega=\pm \Omega$.

The other contributions to the current noise stem from diagrams containing two oscillator lines $D(t,t')$. These diagrams are
all proportional to $\beta_{1} \beta^{*}_{1} \beta_{1} \beta^{*}_{1} = t_{1}^{4}$ and therefore independent of the relative phase
$\eta$ between $t_{0}$ and $t_{1}$. Moreover, these current noise contributions are small compared to the ones containing only one
oscillator line since $t_{1}^{4} \ll t_{0}^{2} \, t_{1}^{2}$. We however are mainly interested in the possibility to detect the oscillator's
momentum which depends on $\eta$, for this reasons and the fact that they are small compared to $\hat{S}^{(4)}_{\rm sym,D}(\omega)$
we do not include them in our discussion, nevertheless we state our result which we obtain after summing up the diagrams
\begin{align}\label{eqn:l99}
	&\hat{S}^{(4)ij}_{DD}(\omega) 	= -\frac{e^{2}}{2 \pi} \int d\omega_{1} \sum_{k,l=\pm} (kl) \Big\{ \nn \\
						& D^{kl}(\omega_{1}) \, D^{ij}(\omega_{1}+\omega) \, \bar{\fg}^{kj}_{11}(-\omega_{1}) \, \bar{\fg}^{li}_{11}(\omega_{1}) + \nn \\
						+& D^{ki}(\omega_{1}) \, D^{lj}(\omega_{1}+\omega) \, \bar{\fg}^{kj}_{11}(-\omega_{1}) \, \bar{\fg}^{li}_{11}(-\omega_{1}-\omega) - \nn \\
						-& D^{ki}(\omega_{1}) \, D^{lj}(-\omega_{1}) \, \fg^{kl}_{11}(-\omega_{1}) \, \fg^{ij}_{11}(\omega_{1}+\omega) \Big\} \, .
\end{align}
The last integration in Eq.~(\ref{eqn:l99}) can be easily done since the oscillator correlation functions are peaked at $\pm \Omega$. Our result is then in accordance with the one obtained by Wabnig {\it et al.} in Ref.~\onlinecite{Wabnig:2007p16}, where it is has been shown that these contributions to the current noise are peaked at $\omega=-2\Omega, 0 , 2\Omega$ in contrast to the contributions arising
from Eq.~(\ref{eqn:l1}). Similar to $\check{S}^{(4)}_{D}(\omega)$, $\check{S}^{(4)}_{DD}(\omega)$ is of the kink-like structure
and therefore only modifying the current noise floor.

We now address the current noise stemming from $\hat{S}^{(4)}_{\rm sym,D}(\omega)$ for arbitrary $\eta$ and also arbitrary
system parameters.

\subsubsection{Current noise in the Markovian and non-Markovian regime for arbitrary $\eta$}

In order to compare our result for the momentum detector with Ref.~\onlinecite{Doiron:2008p13}, we investigate $\hat{S}^{(4)}_{\rm sym,D}(\omega)$
near the resonance ($\omega \approx \Omega$). We find
\begin{align}\label{eqn:m1}
	& \hat{S}^{(4)}_{\rm sym,D}(\omega)	\approx 4 \pi^{2} e^{2} \rho_{0}^{4} \, t_{0}^{2}t_{1}^{2} \, \Bigg\{ \nn \\
								& 4 e^{2} V^{2} \cos(\eta)^{2} \left[1 - \frac{\sigma^{-}(\Omega,V)}{4eVm\Omega \ex{\bar{x}^{2}}} \, \sqrt{1-(\frac{\gamma_{tot}}{\Omega})^{2}} \right] \, S_{x}(\omega) + \nn \\
								+& \frac{1}{m^{2}} \sin(\eta)^{2} \left[1- \frac{2 \sigma^{+}(\Omega,V) m}{\ex{\bar{p}^{2}}} \, \sqrt{1-(\frac{\gamma_{tot}}{\Omega})^{2}}  \right] \, S_{p}(\omega) + \nn \\
								+& \cos(\eta) \sin(\eta) \frac{1}{m} \frac{\omega^{2}-\Omega^{2}}{4 \gamma_{tot}^{2} \Omega^{2} + (\omega^{2}-\Omega^{2})^{2}} \times \nn \\
								& \qquad \times \left[ 4eV \sigma^{+}(\Omega,V) - 2\Omega \sigma^{-}(\Omega,V) \right] \Bigg\} \, ,
\end{align}
where $\sigma^{\pm}(\xi,V)$ is given in Eq.~(\ref{eqn:i2}) and $S_{x}(\omega)$ and $S_{p}(\omega)$ near resonance are given
by
\begin{align}\label{eqn:m2}
	S_{X}(\omega) = \frac{2 \gamma_{\rm tot}^{2} \Omega^{2} \ex{X^{2}}}{4 \gamma_{\rm tot}^{2} \Omega^{2} + (\omega^{2}-\Omega^{2})^{2}} \, ,
\end{align}
with $X=\bar{x},\bar{p}$. The above expression is valid for the Markovian as well as for the non-Markovian regime. The relevant information
about the oscillator can now be gained form the current noise spectrum.

We take two different ways of evaluating the expectation values $\ex{\bar{x}^{2}}$ and $\ex{\bar{p}^{2}}$. In the first one we use
number-states which lead to expectation values $\ex{\bar{x}^{2}}_{n} = (2n+1)/m\Omega$ and $\ex{\bar{p}^{2}}_{n} = (2n+1) \, m \Omega$,
where $n$ denotes the oscillator's number of quanta.

Since we also could imagine, as already explained in Sec.~\ref{sec:oscStat1},  two equilibrium bathes, the tunnel junction and
an external heat bath, to which the oscillator couples, we can assign an effective temperature $T_{\rm eff}$ to the oscillator which
obeys $\gamma_{\rm tot} \, T_{\rm eff} = \gamma_{0} \, T_{\rm env} + \gamma_{+} \, T_{\rm junc}$, where $\gamma_{\rm tot} = \gamma_{0} + \gamma_{+}$
is the total damping due to coupling to the junction ($\gamma_{+}$) and an external heat bath ($\gamma_{0}$). The external
heat bath is at temperature $T_{\rm env}$ and the junction's temperature is given by $T_{\rm junc} = eV/2k_{B}$. The oscillator's
expectation values in this thermal regime are then give by $\ex{\bar{x}^{2}} = 2 \, k_{B} \, T_{\rm eff} / m\, \Omega^{2}$ and
$\ex{\bar{p}^{2}} = 2 \, m \, k_{B} \, T_{\rm eff}$.

For both cases, the thermal case and the number-state one, it is convenient scaling the current noise $\hat{S}_{\rm sym,D}^{(4)}(\omega)$
with $e I_{0} = 2 \pi \rho_{0}^{2} e^{2} t_{0}^{2} \sigma^{+}(\Omega,V)$. We furthermore introduce dimensionless constant
$f_{1},f_{2},f_{3},f_{4}$ which are defined in the following way
\begin{align}
	\gamma_{\rm tot}	&= \frac{\Omega}{f_{1}} \\
	\gamma_{+}		&= \frac{\gamma_{\rm tot}}{f_{2}} = \frac{\Omega}{f_{1} \, f_{2}} \\
	eV				&= f_{3} \, \Omega \\
	T_{\rm env}		&= f_{4} \frac{eV}{k_{B}} \, .
\end{align}
$f_{1}$ can be interpreted as an overall quality-factor. The ratio $\gamma_{0}/\gamma_{+} = (f_{2}-1)$ leads for $f_{2} \in ]1,2[$
to a stronger coupling to the tunnel junction and for $f_{2} > 2$ to a stronger coupling to the external heat bath. The parameter
$f_{3}$ distinguishes the non-Markovian ($f_{3} \in ]0,1] $) from the Markovian regime ($f_{3} \gg 1 $). The last parameter $f_{4}$
quantifies the temperature $T_{\rm env}$ of the external bath wit respect to the applied bias $V$.

We now compare the signal of the position detector $S^{(4)}_{\rm x-det}(\omega) =  \hat{S}^{(4)}_{\rm sym,D}(\omega;\eta=0)$ to the signal
of the momentum detector $S^{(4)}_{\rm p-det}(\omega) =  \hat{S}^{(4)}_{\rm sym,D}(\omega;\eta=\pi/2)$ and later their dependencies on
the parameters $f_{i}$ at resonance $\omega \approx \Omega$. Assuming a high quality resonator we take
$\sqrt{1- \gamma_{\rm tot}^{2}/\Omega^{2}} \approx 1$ in Eq.~(\ref{eqn:m1}). We call $\fq_{x} = \sigma^{-}(\Omega,V) / (4eVm\Omega \ex{\bar{x}^{2}})$
quantum corrections to the $x$-detector current noise, arising from the non-vanishing commutator $[\hat{x},\hat{p}]$, similarly
we call $\fq_{p} = (2 m \sigma^{+}(\Omega,V)) / \ex{\bar{p}^{2}}$ quantum corrections to the $p$-detector current noise. We
then find
\begin{align}
	S^{(4)}_{\rm x-det} = 4 \, f_{3}^{2} \, \frac{[1-\fq_{x}]}{[1-\fq_{p}]} \, S^{(4)}_{\rm p-det}
\end{align}
and conclude that in the Markovian regime where $f_{3} \gg 1$ the signal of the position detector is always larger than the signal of
the momentum detector. Whereas in the non-Markovian regime we have a stronger signature of the noise part showing the
momentum signature of the oscillator. In the following, we investigate in more detail the current noise of the $x$- and $p$-detector.

\subsubsection{The x-detector}

\begin{figure}[ht]
	\center{\includegraphics[width=0.75\columnwidth]{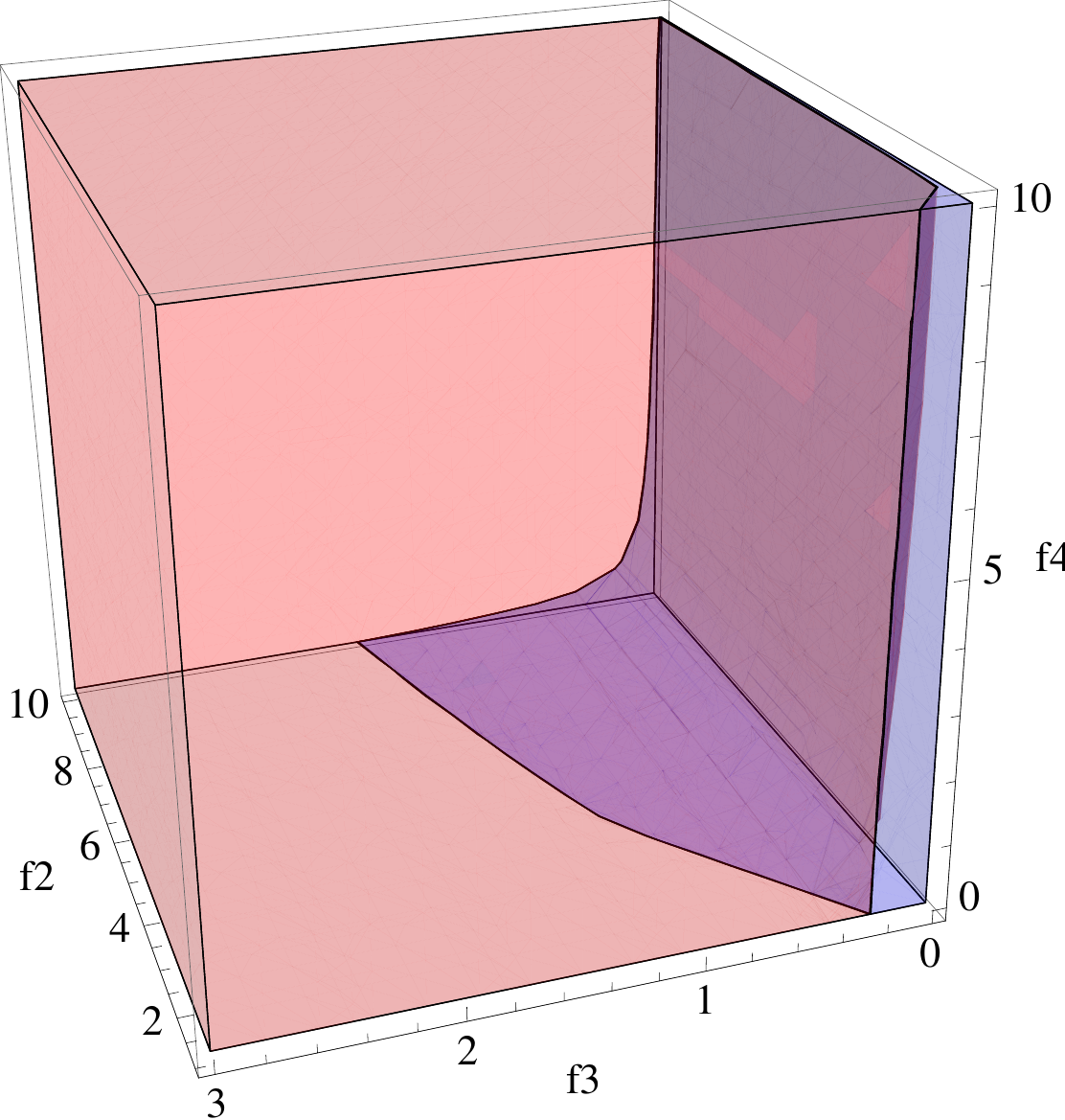}}
	\caption{ The current noise peak at $\omega = \Omega$ for the $x$-detector as a function of $f_{2},f_{3},f_{4}$, where we
			took $f_{1} \rightarrow \infty$.
			The blue region shows the parameter regime where the peak is negative, for parameter combinations lying
			in the red region, the peak is positive. }
	\label{fig:xDet1}
\end{figure}

From Eq.~(\ref{eqn:m1}) one can see that for $\eta=0 \mod \pi$ we recover the position detector result as in Refs.~\onlinecite{
Doiron:2007p5, Clerk:2004p185, Wabnig:2007p16}, with peaks in the current noise spectrum at $\omega=\pm \Omega$.
Since we calculate the symmetrized current-current correlator, the current noise is symmetric in $\omega$. The sign of the
signal is given by the sign of $[1-\fq_{x}]$ which for an oscillator in the thermal regime depends on $f_{2},f_{3}$ and $f_{4}$,
for an oscillator in number-state $n$ it depends on $n$ and $f_{3}$ only. We stick to a thermal resonator, noting that as in
Ref.~\onlinecite{ Doiron:2008p13} for the $p$-detector, here the quantum corrections $\fq_{x} \sim 1/f_{3}$ can become large
(compared to $1$) in the non-Markovian regime where $f_{3} < 1$, leading to a sign change. The parameter regimes for a
negative/positive peak in the current noise are depicted in Fig.~\ref{fig:xDet1} as blue/red regions.
The change of sign in the signal, depends on the environment temperature $T_{\rm env}$, the coupling to the environments $f_{2}$
and heavily on the bias voltage and therefore $f_{3}$. Deep in the Markovian regime the sign change is hard to achieve, only if $f_{2}$
is very large and the bath temperature $T_{\rm env}$ is very low, meaning that heating of the oscillator can be compensated by strongly
coupling to a cold environment. In the non-Markovian regime the quantum corrections $\fq_{x}$ can become large more easily and
due to the lower signal in the non-Markovian regime for the $x$-detector, the sign change is more pronounced.

\begin{figure}[ht]
	\center{\includegraphics[width=1\columnwidth]{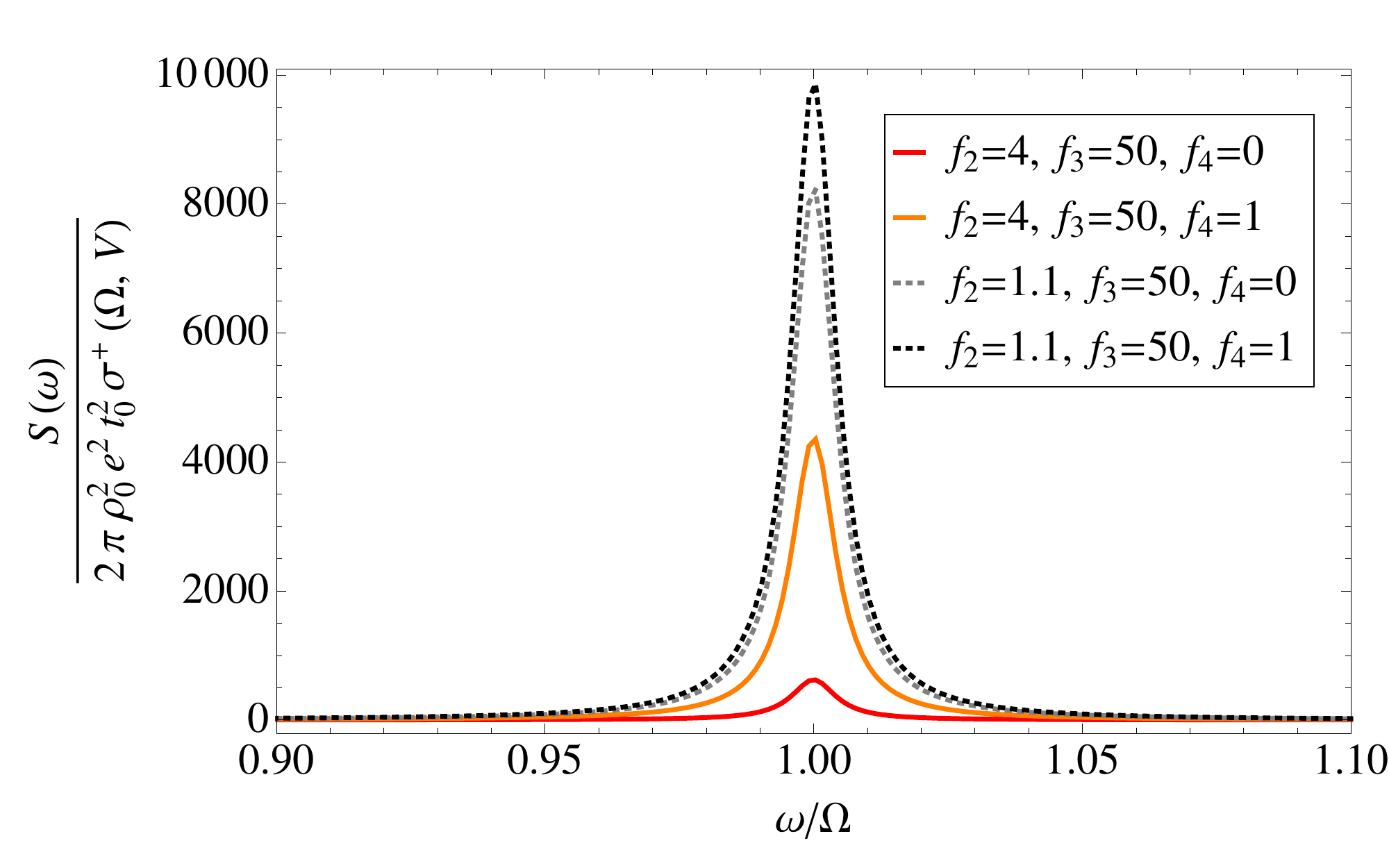}}
	\caption{ Noise signal at $\omega = \Omega$ in the Markovian regime for different values of $f_{2},f_{3},f_{4}$ and $f_{1}=200$
			which shows a peak, due to the presence of the oscillator. }
	\label{fig:xDet2}
\end{figure}

\begin{figure}[ht]
	\center{\includegraphics[width=1\columnwidth]{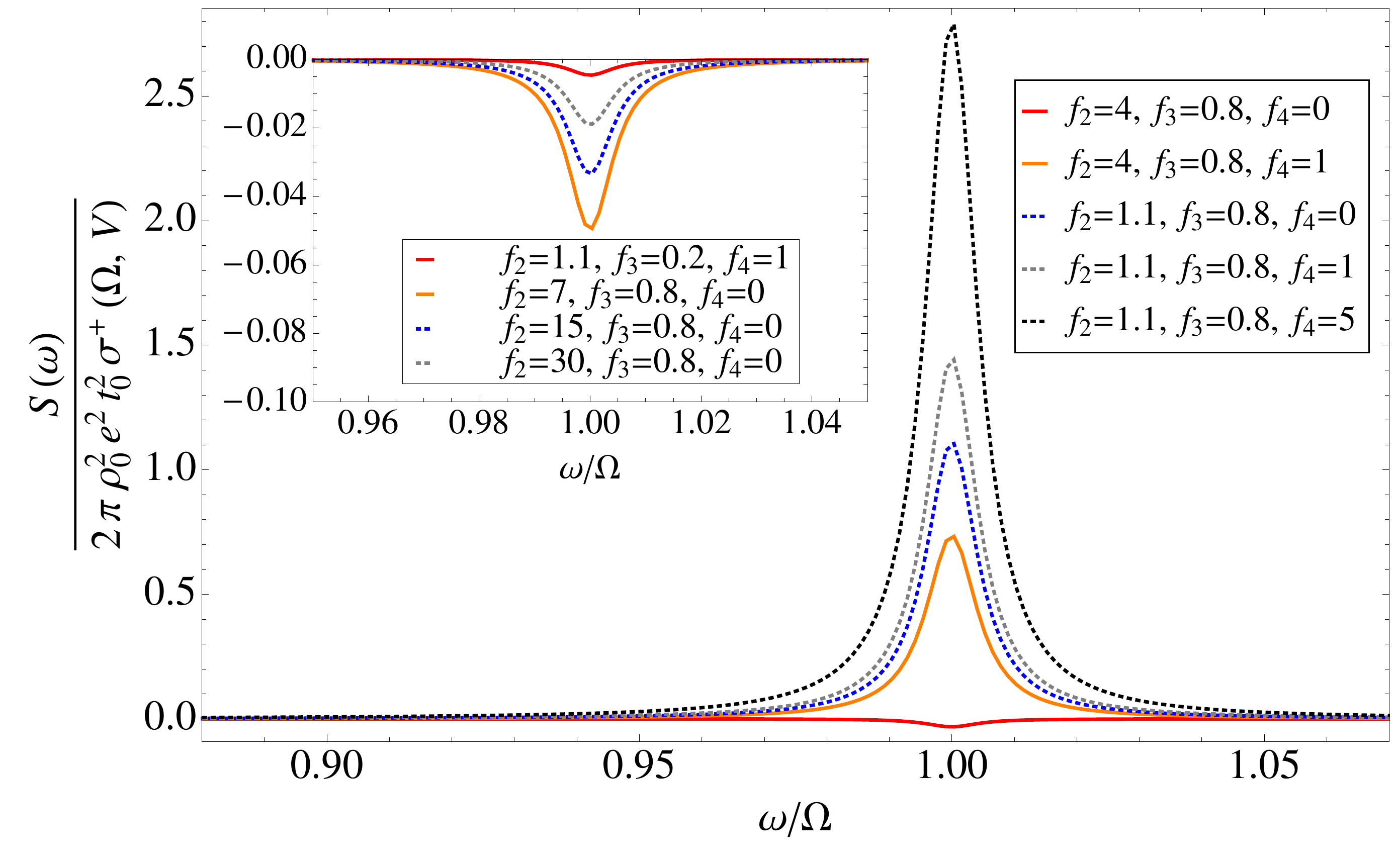}}
	\caption{ $x$-detector current noise signal at $\omega = \Omega$ in the non-Markovian regime for different values of $f_{2},f_{3},f_{4}$ and $f_{1}=200$.
			The signal is weaker than in the Markovian regime (compare to Fig.~\ref{fig:xDet2}). In this regime however, it is possible to see a change of sign of the
			signal, depending on the parameters $f_{2},f_{4}$. The inset illustrates this sign change.  }
	\label{fig:xDet3}
\end{figure}

Figs.~\ref{fig:xDet2} and \ref{fig:xDet3} show the current noise spectrum around $\omega \approx \Omega$ in the Markovian and the non-Markovian
regime for different couplings and environment temperatures.

\subsubsection{The p-detector}
\label{sec_pdet}

\begin{figure}[ht]
	\center{\includegraphics[width=0.75\columnwidth]{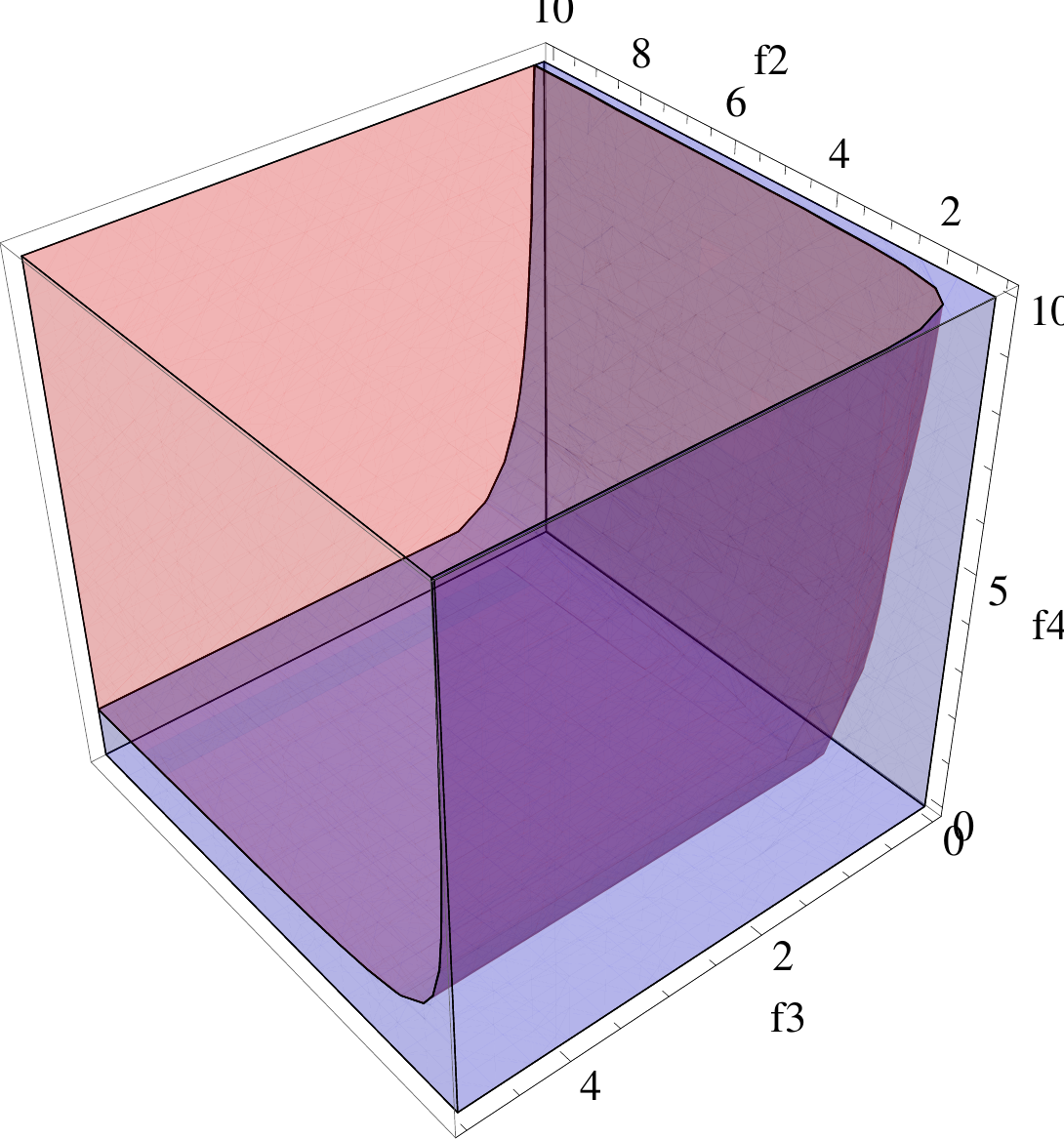}}
	\caption{ This figure shows the parameter regimes for the $p$-detector where the current noise peak at $\omega = \Omega$
			has a negative sign (blue) and positive sign (red). Compared to the $x$-detector the sign change is possible for a
			wider range of parameters $f_{2},f_{3},f_{4}$; we took $f_{1} \rightarrow \infty$. }
	\label{fig:pDet1}
\end{figure}

\begin{figure}[ht]
	\center{\includegraphics[width=1\columnwidth]{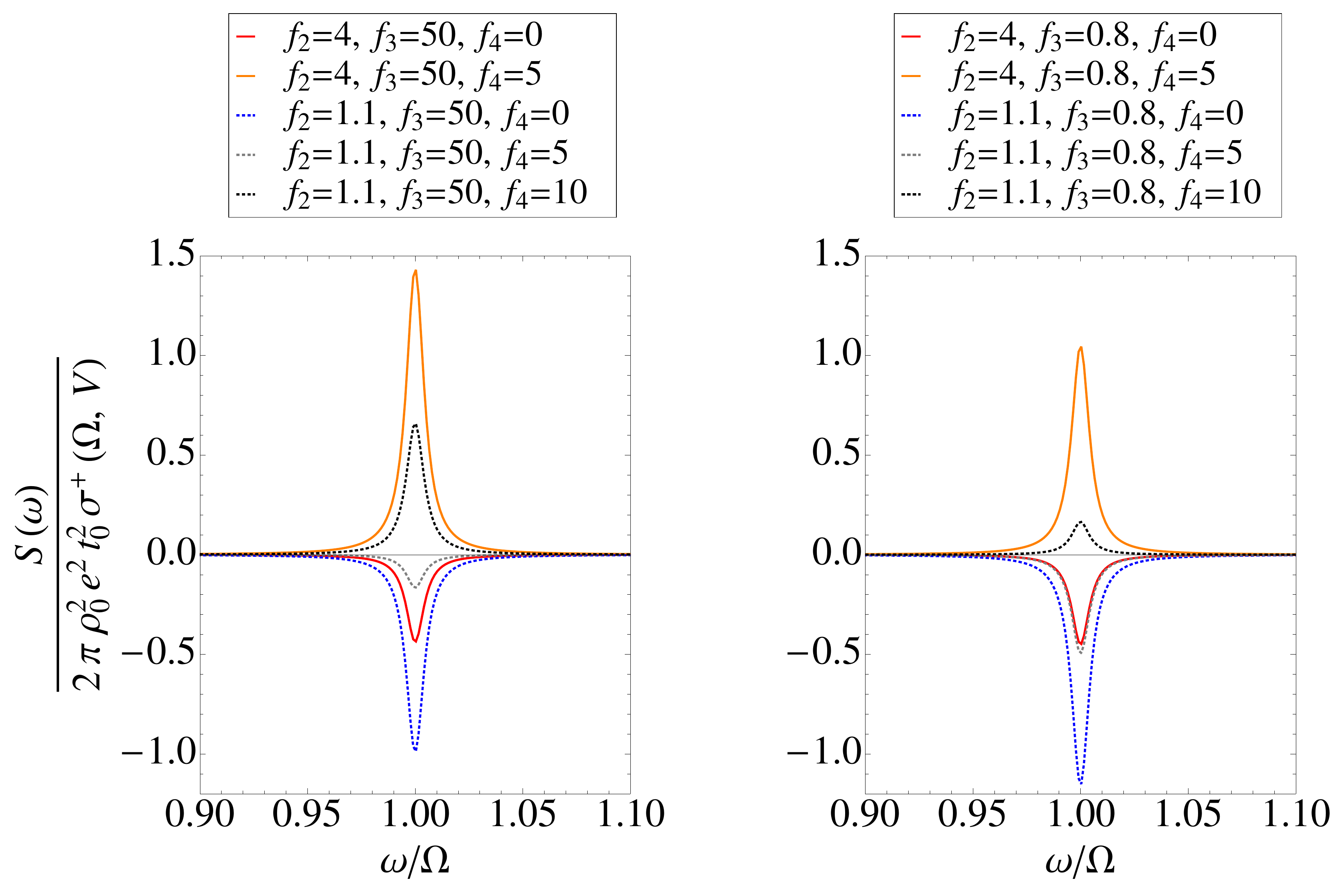}}
	\caption{ The left panel shows the $p$-detector current noise in the Markovian regime for various parameters $f_{2},f_{4}$ and $f_{1}=200$.
			The right panel shows the current noise in the non-Markovian regime for various parameters $f_{2},f_{4}$. In
			both cases we can easily achieve a sign change in the signal. }
	\label{fig:pDet2}
\end{figure}

For the cases $\eta=\pi/2 \mod \pi$ in Eq.~(\ref{eqn:m1}), the result of the momentum detector as stated in Ref.~\onlinecite{
Doiron:2008p13} are extended to the non-Makovian regime. Due to the fact that the quantum corrections $\fq_{p}$ are in the first place larger than $\fq_{x}$,
the peak in the current noise spectrum stemming from the oscillator has a negative sign for $\eta=\pi/2$. However, it is also
possible to change the sign by adjusting the parameters $f_{2},f_{3},f_{4}$ on which $\fq_{p}$ depends. In Fig.~\ref{fig:pDet1}
we depict  the regions with a negative sign blue and the ones with a positive sign red. Changing the sign of the current
noise signal in the $p$-detector case is easier to achieve over a wide range of parameters, as compared to the $x$-detector,
even deep in the Markovian regime ($f_{3} \gg 1$). Figure~\ref{fig:pDet2} shows a summary of the $p$-detector
current noise in the Markovian and non-Markovian regime for different parameters $f_{2}, f_{4}$, respectively.
%

\subsubsection{Detection of number states}
\label{sec_number}

\begin{figure}[ht]
	\center{\includegraphics[width=0.7\columnwidth]{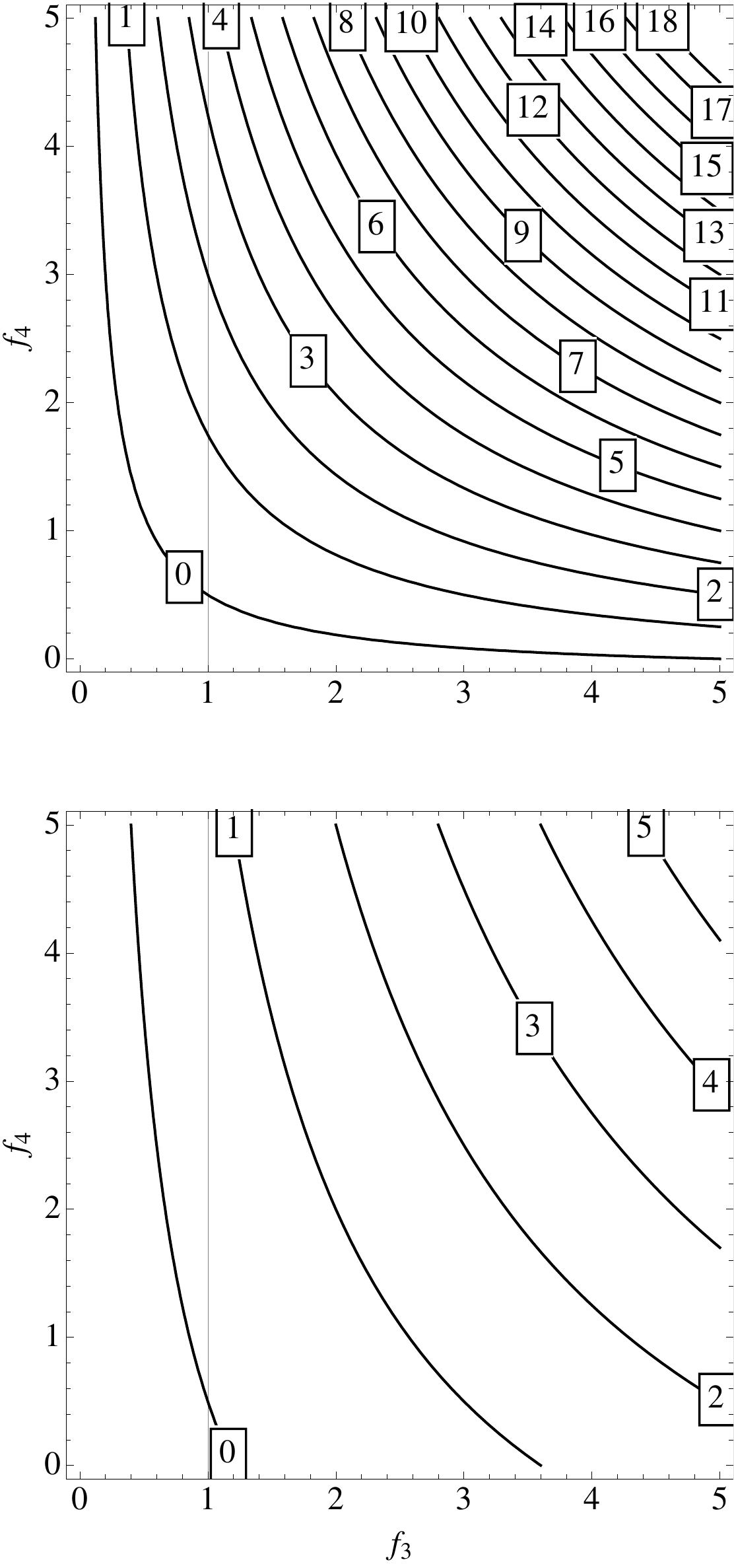}}
	\caption{ Number of quanta $n$ on the oscillator in dependence of experimentally adjustable parameters,
			for dominant coupling to the external heat bath (upper panel, with $f_{2}=5$) and dominant
			coupling to the tunnel junction (lower panel, with $f_{2}=1.2$). }
	\label{fig:nThmap1}
\end{figure}

With the above made definitions we can map the occupation number of the oscillator to experimentally adjustable parameters,
as for instance the environment temperature $T_{\rm env}$ and the bias voltage $V$, similar to the approach in Ref.~\onlinecite{Wabnig:2007p16}.
This allows us in general to determine which state $n$ the oscillator is in. This mapping is independent of $\eta$ and depends
only on the dimensionless parameters $f_{i}$ in the following way
\begin{align}\label{eqn:n1}
	n = \frac{2 f_{2} |f_{3}| f_{4} - 2 |f_{3}| f_{4} - f_{2} + |f_{3}|}{2 f_{2}} \, .
\end{align}
The dependance of $n$ on adjustable parameters is depicted in Fig.~\ref{fig:nThmap1}.
In the Markovian regime the oscillator is only in its ground state for low environmental temperatures, since the applied bias
voltage is heating the oscillator. In the non-Markovian regime, we can have a higher environmental temperature for the
oscillator being in its ground state.

\section{Conclusion}
\label{sec:conclusion}

We have studied the finite frequency current noise of a tunnel junction coupled to a harmonic oscillator. In our work, we go
beyond the Born approximation (because we calculate the noise up to fourth order in the tunneling amplitude) and beyond
the Markov approximation (because we do not restrict ourselves to the regime $eV/\hbar \Omega \gg 1$). For a non-stationary
oscillator, we have shown that the finite frequency current noise of the detector can be complex. This complex current noise
can be used to obtain information about expectation values depending on $\hat{x}$ as well as expectation values depending
on $\hat{p}$. The former we call $x$-detector signal and the latter $p$-detector signal.

For the stationary oscillator, the finite frequency current noise is always real. Then, it is more complicated to get momentum
information using a tunnel junction detector. An Aharonov-Bohm-loop setup is needed for this task. We analyze such a setup
for the first time in the non-Markovian regime and thereby show how the $x$- and the $p$-signal can be used to determine
the quantum state of the oscillator.

Our analysis is an essential prerequisite to study how the quantum entanglement of NEMS \cite{Eisert:2004} can be measured on the basis of tunnel junction detectors. This very interesting problem will be addressed in future work.

\begin{acknowledgments}
	We thank T.~L.~Schmidt and K.~B{\o}rkje for interesting discussions and acknowledge financial support from the DFG.
\end{acknowledgments}

\begin{widetext}

\appendix

\section{Details on the fermionic Green's functions}
\label{sec:appendix1}

Expectation values of the fermionic operators $\hat{h}_{i}$ and $\hat{j}_{i}$ can be expressed by the Keldysh Green's functions
$G_{l,r}(t,t') = -i \ex{T_{c} \, \hat{c}_{l,r}(t) \hat{c}_{l,r}^{\dag}(t')}$ as
\begin{align}\label{eqn:aa11}
	\ex{T_{c} \, \hat{h}_{i}(t) \, \hat{j}_{j}(t') } &= -i \, \bar{\fg}_{ij}(t,t') = -i \left[ \beta_{i} \beta_{j}^{*} G_{r}(t,t') G_{l}(t',t) - \beta_{i}^{*} \beta_{j} G_{r}(t',t) G_{l}(t,t') \right] \\
	\ex{T_{c} \, \hat{h}_{i}(t) \, \hat{h}_{j}(t') } &= \fg_{ij}(t,t') = \left[ \beta_{i} \beta_{j}^{*} G_{r}(t,t') G_{l}(t',t) + \beta_{i}^{*} \beta_{j} G_{r}(t',t) G_{l}(t,t') \right] \\
	\ex{T_{c} \, \hat{j}_{i}(t) \, \hat{j}_{j}(t') } &= \fg_{ij}(t,t') = \left[ \beta_{i} \beta_{j}^{*} G_{r}(t,t') G_{l}(t',t) + \beta_{i}^{*} \beta_{j} G_{r}(t',t) G_{l}(t,t') \right]
\end{align}
The Fourier transform of the function $\fg_{ij}(t,t')$ can be calculated in the usual way, yielding
\begin{align}
	\fg_{ij}^{-+}(\w)^{+} &= 2 \pi \rho_{0}^{2} \left[ \beta_{i}\beta_{j}^{*} \frac{eV+\w}{2} [-1 + \coth(\beta\frac{eV+\w}{2})] + \beta_{i}^{*}\beta_{j} \frac{eV-\w}{2} [1 + \coth(\beta\frac{eV-\w}{2})] \right] \label{eqn:aa12} \\
	\fg_{ij}^{+-}(\w)^{+} &= 2 \pi \rho_{0}^{2} \left[ \beta_{i}\beta_{j}^{*} \frac{eV+\w}{2} [1 + \coth(\beta\frac{eV+\w}{2})] + \beta_{i}^{*}\beta_{j} \frac{eV-\w}{2} [-1 + \coth(\beta\frac{eV-\w}{2})] \right] \label{eqn:aa13}
\end{align}
\begin{align}\label{eqn:aa14}
	\bar{\fg}_{ii}^{--}(\omega) 	&= \bar{\fg}_{ii}^{++}(\omega) = \beta_{i}\beta_{i}^{*} \int \frac{d\omega_{1}}{2\pi} \left[ G_{r}^{--}(\omega_{1}+\omega)G_{l}^{--}(\omega_{1}) - G_{r}^{--}(\omega_{1})G_{l}^{--}(\omega_{1}+\omega)  \right] \nn \\
						&= 2 \, \pi \, \rho_{0}^{2} \, \beta_{i}\beta_{i}^{*} \, \sigma^{-}(\omega,V) \, ,						
\end{align}
where $\sigma^{-}(\omega,V)$ is given in Eq.~(\ref{eqn:i2}).

\section{Details of the second order current noise calculation}
\label{sec:appendix2}

The starting point for the calculation is Eq.~(\ref{eqn:h2}), together with Eqs.~(\ref{eqn:d1}, \ref{eqn:d2}) the symmetrized current noise
in the non-stationary case can be written as
\begin{align}\label{eqn:aa21}
	S^{(2)}_{\rm sym}(\omega,t') &= \frac{e^{2}}{2} \int dt \, e^{i \omega t} \Big\{ \fg_{00}^{-+}(t) + \fg_{00}^{+-}(t) + \nn \\
				+& \ex{\hat{x}(t')} \left[ \fg_{01}^{-+}(t) + \fg_{01}^{+-}(t) \right]
				+ \ex{\hat{x}(t')} \cos(\Omega t) \left[ \fg_{10}^{-+}(t) + \fg_{10}^{-+}(t) \right]
				+ \frac{\ex{\hat{p}(t')}}{m\Omega} \sin(\Omega t) \left[ \fg_{10}^{-+}(t) + \fg_{10}^{+-}(t) \right] + \nn \\
				+& \ex{\hat{x}(t') \hat{x}(t')} \cos(\Omega t) \left[ \fg_{11}^{-+}(t) + \fg_{11}^{+-}(t) \right]
				+ \frac{\ex{\hat{x}(t') \hat{p}(t')}}{m\Omega} \sin(\Omega t) \fg_{11}^{-+}(t)
				+ \frac{\ex{\hat{p}(t') \hat{x}(t')}}{m\Omega} \sin(\Omega t) \fg_{11}^{+-}(t) \Big\} \, .
\end{align}
The further calculation is straightforward by using Eq.~(\ref{eqn:aa12}) and Eq.~(\ref{eqn:aa13}). Finally, the current noise $S^{(2)}_{\rm sym}(\omega,t')$
in Eq.~(\ref{eqn:aa21}) can be written as
\begin{align}\label{eqn:aa22}
	&S^{(2)}_{\rm sym}(\omega,t')= \frac{e^{2}}{2} \Big\{ \fg_{00}^{-+}(\w) + \fg_{00}^{+-}(\w) + \ex{\hat{x}(t')} \left[ \fg_{01}^{-+}(\w) + \fg_{01}^{+-}(\w) \right] + \nn \\
				&+ \frac{1}{2} \ex{\hat{x}(t')} \big[ \fg_{10}^{-+}(\w+\W) + \fg_{10}^{-+}(\w-\W) + \fg_{10}^{+-}(\w+\W) + \fg_{10}^{+-}(\w-\W) \big] - \nn \\
				&- \frac{i}{2 m \Omega} \ex{\hat{p}(t')} \big[ \fg_{10}^{-+}(\w+\W) - \fg_{10}^{-+}(\w-\W) + \fg_{10}^{+-}(\w+\W) - \fg_{10}^{+-}(\w-\W) \big] + \nn \\
				&+ \frac{1}{2} \ex{\hat{x}(t') \hat{x}(t')} \big[ \fg_{11}^{-+}(\w+\W) + \fg_{11}^{-+}(\w-\W) + \fg_{11}^{+-}(\w+\W) + \fg_{11}^{+-}(\w-\W) \big] -\nn \\
				&- \frac{i}{2 m \Omega} \ex{\hat{x}(t') \hat{p}(t')} \left[ \fg_{11}^{-+}(\w+\W) - \fg_{11}^{-+}(\w-\W) \right] - \frac{i}{2 m \Omega} \ex{\hat{p}(t') \hat{x}(t')} \left[ \fg_{11}^{+-}(\w+\W) - \fg_{11}^{+-}(\w-\W) \right] \Big\} \, .
\end{align}
The functions $\sigma^{\pm}(\xi,V)$, see Eq.~(\ref{eqn:i2}), allow us to distinguishing the Markovian from the non-Markovian regime. For $T \rightarrow 0$ we find
\begin{align}
	\sigma^{-}(\xi,V) &= \begin{cases} {\rm sgn}(V) \, \xi & e|V| > \xi \\ {\rm sgn}(V) \, e |V| & e|V| < \xi \end{cases} \\
	\sigma^{+}(\xi,V) &= \begin{cases} e|V| & e|V| > \xi \\ \xi & e|V| < \xi \end{cases} \, ,
\end{align}
where $T$ here is the temperature of electrons in the leads.

\section{Details of the fourth order current noise calculation}
\label{sec:appendix3}

We first give the whole expression for the current noise to fourth order in the tunneling amplitudes containing the $\fm$-functions of
Eq.~(\ref{eqn:k1}) and oscillator operators
\begin{align}\label{eqn:aa31}
	S^{(4)}&(\tau_{3},\tau_{4}) = -\frac{e^{2}}{2} \int_{c}d\tau_{1} \, d\tau_{2} \,\, \Big\{ \nn \\ & \fm_{\rm 0,0,0,0}(\tau_{1},\tau_{2},\tau_{3},\tau_{4}) + \nn \\
	&+ \fm_{\rm 0,0,0,1}(\tau_{1},\tau_{2},\tau_{3},\tau_{4}) \ex{\hat{x}(\tau_{4})} +  \fm_{\rm 0,0,1,0}(\tau_{1},\tau_{2},\tau_{3},\tau_{4}) \ex{\hat{x}(\tau_{3})} + \nn \\
	&+ \fm_{\rm 0,1,0,0}(\tau_{1},\tau_{2},\tau_{3},\tau_{4}) \ex{\hat{x}(\tau_{2})} + \fm_{\rm 1,0,0,0}(\tau_{1},\tau_{2},\tau_{3},\tau_{4}) \ex{\hat{x}(\tau_{1})} + \nn \\
	&+ \fm_{\rm 0,0,1,1}(\tau_{1},\tau_{2},\tau_{3},\tau_{4}) \ex{T_{c} \, \hat{x}(\tau_{3}) \hat{x}(\tau_{4})} + \fm_{\rm 0,1,0,1}(\tau_{1},\tau_{2},\tau_{3},\tau_{4}) \ex{T_{c} \, \hat{x}(\tau_{2}) \hat{x}(\tau_{4})} +\nn \\
	&+ \fm_{\rm 0,1,1,0}(\tau_{1},\tau_{2},\tau_{3},\tau_{4}) \ex{T_{c} \, \hat{x}(\tau_{2}) \hat{x}(\tau_{3})} + \fm_{\rm 1,0,0,1}(\tau_{1},\tau_{2},\tau_{3},\tau_{4}) \ex{T_{c} \, \hat{x}(\tau_{1}) \hat{x}(\tau_{4})} + \nn \\
	&+ \fm_{\rm 1,0,1,0}(\tau_{1},\tau_{2},\tau_{3},\tau_{4}) \ex{T_{c} \, \hat{x}(\tau_{1}) \hat{x}(\tau_{3})} + \fm_{\rm 1,1,0,0}(\tau_{1},\tau_{2},\tau_{3},\tau_{4}) \ex{T_{c} \, \hat{x}(\tau_{1}) \hat{x}(\tau_{2})} + \nn \\
	&+ \fm_{\rm 0,1,1,1}(\tau_{1},\tau_{2},\tau_{3},\tau_{4}) \ex{T_{c} \, \hat{x}(\tau_{2}) \hat{x}(\tau_{3}) \hat{x}(\tau_{4})} + \fm_{\rm 1,0,1,1}(\tau_{1},\tau_{2},\tau_{3},\tau_{4}) \ex{T_{c} \, \hat{x}(\tau_{1}) \hat{x}(\tau_{3}) \hat{x}(\tau_{4})} + \nn \\
	&+ \fm_{\rm 1,1,0,1}(\tau_{1},\tau_{2},\tau_{3},\tau_{4}) \ex{T_{c} \, \hat{x}(\tau_{1}) \hat{x}(\tau_{2}) \hat{x}(\tau_{4})} + \fm_{\rm 1,1,1,0}(\tau_{1},\tau_{2},\tau_{3},\tau_{4}) \ex{T_c \, \hat{x}(\tau_{1}) \hat{x}(\tau_{2}) \hat{x}(\tau_{3})} + \nn \\
	&+ \fm_{\rm 1,1,1,1}(\tau_{1},\tau_{2},\tau_{3},\tau_{4}) \ex{T_{c} \, \hat{x}(\tau_{1}) \hat{x}(\tau_{2}) \hat{x}(\tau_{3}) \hat{x}(\tau_{4})} \Big\}
	+ \int_{c} d\tau_{1} \, d\tau_{2} \ex{ T_{c} \, \hat{H}_{\rm tun}(\tau_{1})\hat{I}(\tau_{3}) } \ex{ T_{c}\hat{H}_{\rm tun}(\tau_{2})\hat{I}(\tau_{4}) } \, .
\end{align}
A first reduction of terms in Eq.~(\ref{eqn:aa31}) is done by only focusing on the stationary case. This allows us to drop terms which
are proportional to $\ex{\hat{x}(t)}$ of Eq.~(\ref{eqn:aa31}) and only keep terms proportional to $D(t,t')$ and $D(t,t') \,D(t'',t''')$.
Unlinked diagrams which appear in this expression are canceled by the $\hat{I}^{2}$ term which is always of the bubble type.

\end{widetext}

\bibliographystyle{apsrev}


\end{document}